\documentclass[journal,10pt,twocolumn,final,]{IEEEtran}
\pdfoutput=1

\usepackage[backend=biber, style=ieee, isbn=false, doi=false]{biblatex}   
\bibliography{references.bib}
\usepackage{booktabs}
\usepackage{tabu}
\usepackage{multirow} 
\usepackage{rotating}

\usepackage[ruled,vlined]{algorithm2e}
\DontPrintSemicolon

\usepackage{mathtools} 
\usepackage{amssymb}   
\usepackage{subdepth} 

\usepackage{amsthm}
\theoremstyle{plain}

\usepackage{empheq}

\usepackage{tikz}

\usepackage{pgfplots}
\usepackage{pgfplotstable}
\usepackage{varwidth}

\usepackage{accents}
\usepackage{placeins}

\usepackage{threeparttable}

\usepackage[caption=false,font=footnotesize]{subfig}

\usepackage{url}
\usepackage{hyperref}

\allowdisplaybreaks

\makeatletter
\newcommand{\removelatexerror}{\let\@latex@error\@gobble}
\makeatother

\usepackage[most]{tcolorbox}
\newtcolorbox{empheqboxed}{colback=white, colframe=black, sharpish corners, top=-2mm, bottom=0pt, left=0mm}

\newlength{\figurewidth}
\newlength{\figureheight}

\usetikzlibrary{arrows}

\definecolor{matlabblue}{rgb}{0 0.4470 0.741}
\definecolor{matlaborange}{rgb}{0.8500 0.3250 0.0980}
\definecolor{matlabyellow}{rgb}{0.9290 0.6940 0.1250}
\definecolor{matlabpurple}{rgb}{0.4940 0.1840 0.5560}
\definecolor{matlabgreen}{rgb}{0.4660 0.6740 0.1880}
\definecolor{matlablightblue}{rgb}{0.3010 0.7450 0.9330}
\definecolor{matlabred}{rgb}{0.6350 0.0780 0.1840}

\definecolor{changeColor}{rgb}{255,0,0}

\pgfplotsset{compat=newest}

\pgfplotscreateplotcyclelist{matlabcolor}{%
	matlabblue, mark=diamond*\\%
	matlaborange, mark=*\\%
	matlabyellow, mark=square*\\%
	matlabpurple, mark=triangle*\\%
	matlabgreen, mark=star\\%
	matlablightblue\\%
	matlabred\\%
}

\pgfplotscreateplotcyclelist{matlabcolorNOmark}{%
	matlabblue\\%
	matlaborange\\%
	matlabyellow\\%
	matlabpurple\\%
	matlabgreen\\%
	matlablightblue\\%
	matlabred\\%
}

\pgfplotsset{
	colormap={parula}{
		rgb255(0)=(249,251,14)
		rgb255(10)=(252,206,46)
		rgb255(15)=(225,185,82)
		rgb255(20)=(165,190,107)
		rgb255(25)=(89,189,140)
		rgb255(30)=(21,177,180)
		rgb255(60)=(7,156,207)
		rgb255(150)=(18,125,216)
		rgb255(300)=(15,92,221)
		rgb255(400)=(53,42,135)
	},
}

\newcommand{\plotfileNOlegend}[1]{
	\pgfplotstableread[col sep=tab]{#1}{\table}
	\pgfplotstablegetcolsof{\table}
	\pgfmathtruncatemacro\numberofcols{\pgfplotsretval-1}
	\pgfplotsinvokeforeach{1,...,\numberofcols}{
		\pgfplotstablegetcolumnnamebyindex{##1}\of{\table}\to{\colname}
		\addplot+[thick,line width=1.5pt] table [y index=##1, col sep=tab] {#1}; 
	}
}

\newcommand{\xn}{\boldsymbol{x}_{n}}

\newcommand{\xinX}{\boldsymbol{x}_{n} \in \mathcal{X}_{m}}
\newcommand{\that}{\skew{1}\hat{t}}

\newcommand{\ba}{\boldsymbol{a}}
\newcommand{\bb}{\boldsymbol{b}}
\newcommand{\bc}{\boldsymbol{c}}
\newcommand{\bx}{\boldsymbol{x}}
\newcommand{\bz}{\boldsymbol{z}}
\newcommand{\bA}{\boldsymbol{A}}
\newcommand{\bB}{\boldsymbol{B}}

\newcommand{\bD}{\boldsymbol{D}}
\newcommand{\bF}{\boldsymbol{F}}

\newcommand{\bI}{\boldsymbol{I}}

\newcommand{\bK}{\boldsymbol{K}}
\newcommand{\bS}{\boldsymbol{S}}

\newcommand{\bX}{\boldsymbol{X}}

\newcommand{\bSigma}{\boldsymbol{\Sigma}}
\newcommand{\btheta}{\boldsymbol{\theta}}
\newcommand{\bTheta}{\boldsymbol{\Theta}}
\newcommand{\bmu}{\boldsymbol{\mu}}
\newcommand{\bPhi}{\boldsymbol{\Phi}}

\newcommand{\bgamma}{\boldsymbol{\gamma}}
\newcommand{\bPsi}{\boldsymbol{\Psi}}
\newcommand{\blambda}{\boldsymbol{\lambda}}
\newcommand{\bOmega}{\boldsymbol{\Omega}}
\newcommand{\bxi}{\boldsymbol{\xi}}

\newcommand{\bxhat}{\hat{\boldsymbol{x}}}
\newcommand{\bxt}{\tilde{\boldsymbol{x}}}
\newcommand{\bxihat}{\skew{2}\hat{\boldsymbol{\xi}}}

\newcommand{\bShat}{\skew{3}\hat{\boldsymbol{S}}}

\newcommand{\bthetahat}{\skew{3}\hat{\btheta}}

\newcommand{\bPhihat}{\hat{\boldsymbol{\Phi}}}

\newcommand{\blambdahat}{\skew{-2}\hat{\boldsymbol{\lambda}}}
\newcommand{\bOmegahat}{\hat{\boldsymbol{\Omega}}}

\DeclareMathOperator*{\argmax}{arg\,max}

\DeclareMathOperator{\erf}{erf}
\DeclareMathOperator{\Tr}{Tr}

\newcommand{\vecop}{\text{vec}}
\newcommand{\vechop}{\text{vech}}

\begin{document}

\title{Real Elliptically Skewed Distributions and Their Application to Robust Cluster Analysis}

\author{Christian~A.~Schroth and Michael Muma,~\IEEEmembership{Member,~IEEE}
\thanks{C. A. Schroth and M. Muma are with the Signal Processing Group at Technische Universit\"at Darmstadt, Germany. mail: \{schroth, muma\}@spg.tu-darmstadt.de}
\thanks{Manuscript submitted, June 29, 2020.}}

\markboth{Submitted to IEEE Transactions on Signal Processing (Accepted)}%
{Schroth \MakeLowercase{\textit{et al.}}: Real Elliptically Skewed Distributions and Their Application to Robust Cluster Analysis}

\maketitle

\begin{abstract}
This article proposes a new class of Real Elliptically Skewed (RESK) distributions and associated clustering algorithms that integrate robustness and skewness into a single unified cluster analysis framework. Non-symmetrically distributed and heavy-tailed data clusters have been reported in a variety of real-world applications. Robustness is essential because a few outlying observations can severely obscure the cluster structure. The RESK distributions are a generalization of the Real Elliptically Symmetric (RES) distributions. To estimate the cluster parameters and memberships, we derive an expectation maximization (EM) algorithm for arbitrary RESK distributions. Special attention is given to a new robust skew-Huber \mbox{M-estimator}, which is also the approximate maximum likelihood estimator (MLE) for the skew-Huber distribution, that belongs to the RESK class. Numerical experiments on simulated and real-world data confirm the usefulness of the proposed methods for skewed and heavy-tailed data sets. 
\end{abstract}

\begin{IEEEkeywords}
robust cluster analysis, real elliptically skewed distributions, EM algorithm, heavy-tailed mixture models, multivariate RES distributions, M-estimator, robust data science, unsupervised learning, skew-Gaussian, skew-t, skew-Huber
\end{IEEEkeywords}

\IEEEpeerreviewmaketitle

\section{Introduction}
\IEEEPARstart{F}{inite} mixture models have been extensively used in cluster analysis to estimate the probability density function (pdf) of a given data set \cite{Dempster.1977, Lloyd.1982, Jain.1988, Arthur.2007, Xu.2015}. It is well-known that estimators that have been derived under a Gaussian data assumption are strongly affected by outliers, which frequently occur in real-world applications \cite{Zoubir.2012, Zoubir.2018, Kadioglu.2018}. A popular remedy is to choose a heavy-tailed finite mixture model, such as, a member of the real elliptically symmetric (RES) family of distributions. The associated maximum-likelihood-estimators (MLE) can handle outliers and heavy-tailed noise. M-estimators provide further robustness, as they decouple the estimation from a specific distribution. The underlying idea is to replace the negative log-likelihood function by a robustness-inducing objective function \cite{Huber.2011, Ollila.2014, Maronna.2018, Liu.2019}. 

While the RES distributions offer a high degree of flexibility in modeling different types of heavy-tailedness in the data distribution, they still depend on the assumption that the data is symmetrically distributed. However, in medical \cite{Wang.2009, Pyne.2009}, localization \cite{Ciosas.2016, Youn.2019} or financial \cite{Bernardi.2012} applications, it is often the case that the data is heavy-tailed \emph{and} non-symmetrically distributed. The focus of this work, therefore, lies on mixture models that account for such cases. The proposed robust clustering algorithms are designed to provide reliable results, even if the data distribution is skewed, heavy-tailed, and if the data set contains outliers. A widely used estimation algorithm for mixture models is the expectation maximization (EM) algorithm \cite{Dempster.1977}. The standard EM algorithm is based on a Gaussian assumption \cite{Bishop.2009}, hence, it is not robust against outliers. Some efforts were made to robustify the EM by outlier detection and removal \cite{Wang.2018, Neykov.2007, Gallegos.2009, Gallegos.2010}, assuming t-distributed/heavy-tailed data \cite{Greselin.2010, Andrews.2012, McNicholas.2012}, using an additional component in a mixture model \cite{Fraley.1998, Dasgupta.1998}, including contaminated Gaussian distributions \cite{Punzo.2013} or robustifying the estimation of the cluster centers and covariances \cite{Roizman.2019}. Recently, an EM algorithm has been derived that is applicable to any RES distribution \cite{Schroth.2021}. An important member of the RES family, is the so-called Huber distribution, for which the popular Huber M-estimator is the MLE.

Multivariate skewed distributions include the skew-Gaussian distribution \cite{Azzalini.1996, Azzalini.2005}, the skew-t distribution \cite{Azzalini.2003} and skew-Laplace distribution \cite{Kotz.2001}. A summary of different skewed distributions can be found in \cite{Genton.2004}, \cite{Azzalini.2013}. In contrast to RES distributions, there are multiple formulations for skewed distributions and, generally speaking, none of the specific formulations seems to be preferable over the others \cite{Lee.2013, Azzalini.2014}. In this work, we use the formulation introduced by \cite{ArellanoValle.2005} because it allows for deriving an EM algorithm with explicit solutions for the model parameters. This does not seem to be possible, for example, for the generalized multivariate skewed distributions introduced in \cite{Branco.2001, Sahu.2003}. 

EM algorithms are available for a wide range of specific skewed distributions, e.g., for the skew-Gaussian \cite{Lin.2009, Vrbik.2012, Cabral.2012, Vrbik.2014b, Vrbik.2014, Pyne.2009b, ArellanoValle.2005}, the skew-t \cite{Lin.2010, Lee.2014, Vrbik.2014b, Vrbik.2014}, the skew power exponential \cite{Dang.2019} and the generalized hyperbolic \cite{Browne.2015}. Nevertheless to the best of our knowledge, a unified estimation framework for a broad class of skewed distributions has not yet been developed. 

To fit a finite mixture model to a given data set, the EM algorithm requires the number of clusters to be known, or estimated. A popular strategy is to use model selection criteria, such as, the Bayesian Information Criterion (BIC) derived by Schwarz \cite{Schwarz.1978, Cavanaugh.1999}. Alternatively, a Bayesian cluster enumeration approach that selects the model that is maximum \textit{a posteriori} most probable \cite{Djuric.1998, Stoica.2004} may be beneficial, because it accounts for the structure of the clustering problem in the penalty term.

Based on \cite{Vrbik.2014b, Vrbik.2014, Pyne.2009b, ArellanoValle.2005}, our main contribution is to propose a formulation for  Real Elliptically Skewed (RESK) distributions and to derive an EM algorithm for such distributions. The RESK family of distributions is a generalization of the RES family of distributions, which are included within the RESK class by choosing an appropriate value for the skewness factor. For robustness purposes, our framework is then generalized to include asymmetric M-estimators, where special attention is given to a new skew-Huber's M-estimator. Numerical experiments demonstrate increased parameter estimation and cluster enumeration performance for skewed and heavy-tailed data. 

The paper is organized as follows. Section~\ref{sec:res_dis} briefly revisits RES distributions and loss functions. Section~\ref{ch:pre_skew} introduces the proposed RESK family of distributions with special attention given to the skew-Huber distribution. Section~\ref{ap:EM_skew} derives the proposed EM algorithm for RESK distributions and Section~\ref{sec:clu_enum} briefly describes cluster enumeration for RESK distributions based on the Schwarz BIC. Simulations and real-world examples are provided and discussed in Section~\ref{sec:sim}. Finally, conclusions are drawn and some possible future steps are mentioned in Section~\ref{sec:conclusion}. 

\textbf{Notation:} Normal-font letters ($a, A$) denote a scalar, bold lowercase ($\ba$) a vector and bold uppercase ($\bA$) a matrix; calligraphic letters ($\mathcal{A}$) denote a set, with the exception of $\mathcal{L}$, which is used for the likelihood function; $\mathbb{R}$ is the set of real numbers and $\mathbb{R}^{r \times 1}$, $\mathbb{R}^{r \times r}$ the set of column vectors of size $r \times 1$ and matrices of size $r \times r$, respectively; $\bA^{-1}$ is the matrix inverse; $\bA^{\top}$ is the matrix transpose; $|a|$ is the absolute value of a scalar; $|\bA|$ is the determinant of a matrix; $\otimes$ represents the Kronecker product; $\vecop(\cdot)$ is the vectorization operator, $\bD$ is the duplication matrix and $\vechop(\cdot)$ is the vector half operator as defined in \cite{Magnus.2007}. The $\vechop$ operator takes a symmetric  $r \times r$ matrix and stacks the lower/upper triangular half into a single vector of length $\frac{r}{2}(r+1)$.
\section{RES Distributions \& Loss Functions}
\label{sec:res_dis}
\subsection{RES Distributions}
Assuming that the observed data $\bx \in \mathbb{R}^{r \times 1}$ follows a RES distribution with centroid $\bmu  \in \mathbb{R}^{r \times 1}$ and positive definite symmetric scatter matrix $\bS \in \mathbb{R}^{r \times r}$, the pdf of $\bx$ is given by
\begin{equation}
f(\bx| \bmu, \bS, g) = \left| \bS \right|^{-\frac{1}{2}} g\left( t\right),
\label{eqn:res}
\end{equation}
where \mbox{$t = \left(\bx - \bmu\right)^{\top} \bS^{-1} \left(\bx - \bmu\right)$} is the squared Mahalanobis distance, see \cite[p.~109]{Zoubir.2018} and \cite{Sahu.2003}. The function $g$, often referred to as the density generator, is defined by
\begin{align}
g(t) = \frac{\Gamma\left(\frac{r}{2}\right)}{\pi^{r/2}} \left(\int_{0}^{\infty} u^{r/2-1} h(u; r) \text{d}u\right)^{-1} h(t; r),
\end{align}
where $ h(t; r)$ is a function for which
\begin{align}
\int_{0}^{\infty} u^{r/2-1} h(u; r) \text{d}u < \infty
\end{align}
holds. Note that $h(t; r)$ can be a function of multiple parameters, not only of $r$. 

\subsection{Loss Functions}
Assuming an observation of $N$ independent and identically distributed (iid) random variables denoted by \mbox{$\mathcal{X} = \left\{\bx_{1}, \dots, \bx_{N}\right\}$}, the associated likelihood function is given by
\begin{equation}
\mathcal{L}(\bmu, \bS|\mathcal{X}) = \prod_{n = 1}^{N}\left| \bS^{-1} \right|^{\frac{1}{2}} g\left( t_{n}\right)
\end{equation}
with $t_{n} = \left(\bx_{n} - \bmu\right)^{\top} \bS^{-1} \left(\bx_{n} - \bmu\right)$. The MLE minimizes the negative log-likelihood function
\begin{align}
-\ln\left(\mathcal{L}(\bmu, \bS|\mathcal{X}) \right) =& -\ln\left( \prod_{n = 1}^{N}\left| \bS^{-1} \right|^{\frac{1}{2}} g\left( t_{n}\right)\right)\notag\\
=& \sum_{n = 1}^{N} -\ln\left(g(t_{n})\right) - \frac{N}{2} \ln\left(\left| \bS^{-1} \right|\right) \notag\\
=& \sum_{n = 1}^{N} \rho_{\text{ML}}(t_{n}) + \frac{N}{2} \ln\left(\left| \bS \right|\right)
\label{eqn:res_like}
\end{align}	
where the ML loss function \cite[p.~109]{Zoubir.2018} is defined as
\begin{equation}
\label{eq:ml-loss}
\rho_{\text{ML}}(t_{n}) = -\ln\left(g(t_{n})\right).
\end{equation}
The corresponding first and second derivatives are denoted, respectively, by
\begin{equation}
\psi_{\text{ML}}(t_{n}) = \frac{\partial \rho_{\text{ML}}(t_{n})}{\partial t_{n}}, \quad \eta_{\text{ML}}(t_{n}) = \frac{\partial \psi_{\text{ML}}(t_{n})}{\partial t_{n}}.
\label{eqn:psi}
\end{equation}
The key idea of M-estimation \cite{Huber.2011} is to replace the ML loss function $\rho_{\text{ML}}(t_{n})$ in Eq.~\eqref{eq:ml-loss} with a more general loss function $\rho(t_{n})$. The aim of robust M-estimators is not necessarily to be optimal for a single specific distribution, but to provide near-optimal performance within a neighborhood of distributions including heavy-tailed distributions. The weight function $\psi(t_{n})$ can be designed according to desired robustness properties. For example, the very popular Huber weight function down weights outlying data points while giving full weight to points with a small Mahalanobis distance. Interestingly, Huber's estimator is not only a popular M-estimator, but it also has been shown to be the MLE for a specific RES distribution. An overview of some exemplary loss functions and their derivatives can be found in Tables~\ref{tb:g_rho_psi} and \ref{tb:eta_PSI}. Since the Gaussian and t distribution are well-known they are not further discussed, but for the Huber distribution a brief discussion is provided in Section~\ref{sec:skew_huber}.

\section{RESK Distributions}
\label{ch:pre_skew}
\subsection{Definition of RESK Distributions}
This section proposes the class of RESK distributions as a generalization of RES distributions to allow for skewness. Well-known skewed distributions, such as, the skew-Gaussian and the skew-t \cite{Sahu.2003, Pyne.2009, Pyne.2009b, Vrbik.2014, Vrbik.2014b} distribution are included as special cases. The aim is to provide a unified framework to derive robust clustering algorithms for skewed data. Based on the RESK distributions, the user is given a large flexibility in designing algorithms that fit to skewed and heavy-tailed data. This is because the proposed formulation allows for the derivation of an explicit solution for the EM algorithm for arbitrary RESK distributions. As an example, a new clustering algorithm by skewing Huber's distribution is derived.

\begin{figure}[t]
	\centering
	\subfloat[$\xi = 3$, $S = 2$] {\resizebox{0.48\columnwidth}{!}{\begin{tikzpicture}
\begin{axis}[
	scaled ticks=false,
	tick label style={/pgf/number format/.cd},
	width = \figurewidth,
	height = \figureheight,
	xmin = -5,
	xmax = 20,
	ymax = 0.3,
	ymin = 0,
	ytick={0,0.05,0.1,0.15,0.2,0.25,0.3},
	yticklabels={$0$,$0.05$,$0.1$,$0.15$,$0.2$,$0.25$,$0.3$},
	grid,
	xlabel= $x$,
	ylabel= pdf,
	legend entries={{$\lambda = -2$}\\
		{$\lambda = 0$}\\
		{$\lambda = 1$}\\
		{$\lambda = 5$}\\
		{Gauß}\\},
	legend pos=north east,
	legend cell align={left},
	]
	\addplot+[thick,matlabblue,no marks,line width=1.5pt] table[x index=0,y index=1,col sep=tab]  {figures/pdf_gaus_lambda_xi3_S2.csv};
	\addplot+[thick,matlaborange,no marks,line width=1.5pt] table[x index=0,y index=2,col sep=tab] {figures/pdf_gaus_lambda_xi3_S2.csv};
	\addplot+[thick,matlabyellow,no marks,line width=1.5pt] table[x index=0,y index=3,col sep=tab] {figures/pdf_gaus_lambda_xi3_S2.csv};
	\addplot+[thick,matlabpurple,no marks,line width=1.5pt] table[x index=0,y index=4,col sep=tab] {figures/pdf_gaus_lambda_xi3_S2.csv};
	\addplot+[thick,black,no marks,dashed, line width=1.5pt] table[x index=0,y index=2,col sep=tab] {figures/pdf_gaus_lambda_xi3_S2.csv};

	%
\end{axis}
\end{tikzpicture}}%
		\label{fig:pdf_gaus_lambda}}
	\hfil
	\subfloat[$S = 2$] {\resizebox{0.48\columnwidth}{!}{\begin{tikzpicture}
\begin{axis}[
	scaled ticks=false,
	tick label style={/pgf/number format/.cd},
	width = \figurewidth,
	height = \figureheight,
	xmin = -10,
	xmax = 20,
	ymax = 0.3,
	ymin = 0,
	ytick={0,0.05,0.1,0.15,0.2,0.25,0.3},
	yticklabels={$0$,$0.05$,$0.1$,$0.15$,$0.2$,$0.25$,$0.3$},
	grid,
	xlabel= $x$,
	ylabel= pdf,
	legend entries={{$\lambda = -2$, $\xi = -3$}\\
		{$\lambda = -2$, $\xi = 0$}\\
		{$\lambda = 5$, $\xi = 1$}\\
		{$\lambda = 5$, $\xi = 5$}\\},
	legend pos=north east,
	legend cell align={left},
	]
	\addplot+[thick,matlabblue,no marks,line width=1.5pt] table[x index=0,y index=1,col sep=tab]  {figures/pdf_gaus_lambda_xi.csv};
	\addplot+[thick,matlaborange,no marks,line width=1.5pt] table[x index=0,y index=2,col sep=tab] {figures/pdf_gaus_lambda_xi.csv};
	\addplot+[thick,matlabyellow,no marks,line width=1.5pt] table[x index=0,y index=3,col sep=tab] {figures/pdf_gaus_lambda_xi.csv};
	\addplot+[thick,matlabpurple,no marks,line width=1.5pt] table[x index=0,y index=4,col sep=tab] {figures/pdf_gaus_lambda_xi.csv};
	%
\end{axis}
\end{tikzpicture}}%
		\label{fig:pdf_gaus_lambda_xi}}\\
	\subfloat[$\xi = 3$, $S = 2$, $\lambda = 2$] {\resizebox{0.48\columnwidth}{!}{\begin{tikzpicture}
\begin{axis}[
	scaled ticks=false,
	tick label style={/pgf/number format/.cd},
	width = \figurewidth,
	height = \figureheight,
	xmin = -5,
	xmax = 15,
	ymax = 0.25,
	ymin = 0,
	ytick={0,0.05,0.1,0.15,0.2,0.25},
	yticklabels={$0$,$0.05$,$0.1$,$0.15$,$0.2$,$0.25$},
	grid,
	xlabel= $x$,
	ylabel= pdf,
	legend entries={{$q_{H} = 0.71$}\\
		{$q_{H} = 0.99$}\\
		{skew-Gauß}\\},
	legend pos=north east,
	legend cell align={left},
	]
	\addplot+[thick,matlabblue,no marks,line width=1.5pt] table[x index=0,y index=1,col sep=tab]  {figures/pdf_huber_q_xi3_L2_S2.csv};
	\addplot+[thick,matlabyellow,no marks,line width=1.5pt] table[x index=0,y index=3,col sep=tab] {figures/pdf_huber_q_xi3_L2_S2.csv};
	\addplot+[thick,black,no marks,dashed, line width=1.5pt] table[x index=0,y index=1,col sep=tab] {figures/pdf_gaus_xi3_L2_S2.csv};

	%
\end{axis}
\end{tikzpicture}}%
		\label{fig:pdf_huber_q}}
	\hfil	
	\subfloat[$\xi = 0$, $S = 1$, $\lambda = 0$] {\resizebox{0.47\columnwidth}{!}{\begin{tikzpicture}
\begin{axis}[
	scaled ticks=false,
	tick label style={/pgf/number format/.cd},
	width = \figurewidth,
	height = \figureheight,
	xmin = 0,
	xmax = 20,
	ymax = 1,
	ymin = 0,
	grid,
	xlabel= $\underline{t}$,
	ylabel= $\psi(\underline{t})$,
	legend entries={{Gauß}\\
		{t, $\nu = 3$}\\
		{t, $\nu = 9$}\\
		{$q_{H} = 0.8$}\\
		{$q_{H} = 0.99$}\\},
	legend pos=north east,
	legend cell align={left},
	]
	\addplot+[thick,matlabblue,no marks,line width=1.5pt] table[x index=0,y index=1,col sep=tab]  {figures/psi_gaus_t_huber.csv};
	\addplot+[thick,matlaborange,no marks,line width=1.5pt] table[x index=0,y index=2,col sep=tab] {figures/psi_gaus_t_huber.csv};
	\addplot+[thick,matlabyellow,no marks,line width=1.5pt] table[x index=0,y index=3,col sep=tab] {figures/psi_gaus_t_huber.csv};
	\addplot+[thick,matlabpurple,no marks,line width=1.5pt] table[x index=0,y index=4,col sep=tab] {figures/psi_gaus_t_huber.csv};
	\addplot+[thick,matlabgreen,no marks,line width=1.5pt] table[x index=0,y index=5,col sep=tab] {figures/psi_gaus_t_huber.csv};

	%
\end{axis}
\end{tikzpicture}}%
		\label{fig:psi_gaus_t_huber}}
	\caption{The upper left figure shows some exemplary skew-Gaussian pdfs for different values of $\lambda$. For $\lambda = 0$, the skew-Gaussian distribution is equal to the Gaussian distribution. In the upper right figure, the influence of $\xi$ is examined, as it shifts the pdf to the left or right. In the lower left figure, the skew-Huber distribution is shown for different values of $q_{H}$ from Eq. \eqref{eqn:qH}. For $q_{H} \rightarrow 1$ the skew-Huber distribution becomes more and more equal to the skew-Gaussian distribution, while smaller values of $q_{H}$ result in heavier tails. A similar behavior can be observed in the lower right figure, that shows the weight function $\psi(\underline{t})$ for different distributions and different values of $q_{H}$. }
\end{figure}
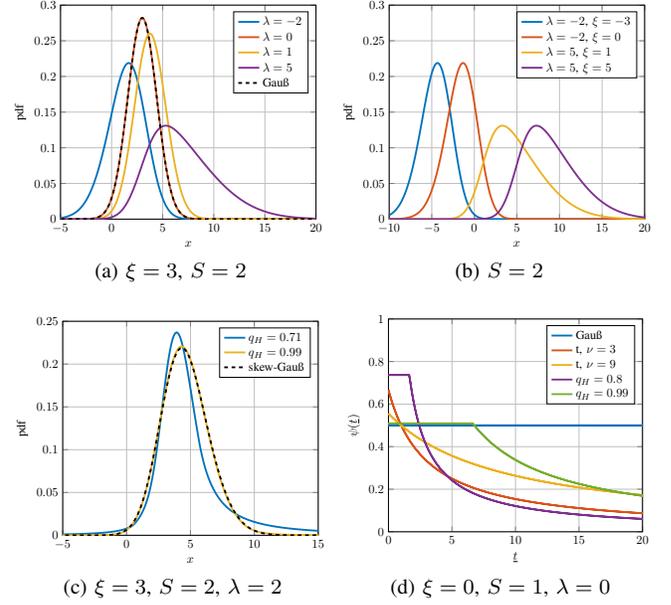

Assuming that the observed data $\bx \in \mathbb{R}^{r \times 1}$ follows a RESK distribution, let $\bxi  \in \mathbb{R}^{r \times 1}$ be the 'centroid', $\blambda \in \mathbb{R}^{r \times 1}$ the skewness factor and let $\bS \in \mathbb{R}^{r \times r}$ be the positive definite symmetric scatter matrix. Then, it is possible to define the RESK distribution as a RES distribution multiplied with its univariate cdf
\begin{equation}
f_{\text{s}}(\bx | \bxi, \bS, \blambda, g) = 2  \left| \bOmega \right|^{-\frac{1}{2}} g\left(\underline{t}\right) F\left(\kappa\left(\underline{t}\right)\right)
\label{eqn:skew_RES}
\end{equation}
with the skewed scatter matrix
\begin{equation}
\bOmega = \bS + \blambda\blambda^{\top},
\label{eqn:skew_cov}
\end{equation}
the skewed squared Mahalanobis distance
\begin{equation}
\underline{t} = \left(\bx - \bxi\right)^{\top} \bOmega^{-1} \left(\bx - \bxi\right)
\label{eqn:skew_maha}
\end{equation}
and
\begin{equation}
\kappa\left(\underline{t}\right) = \frac{\eta}{\tau} \sqrt{2 \psi\left(\underline{t}\right)}
\end{equation}
with $\psi\left(t\right)$ from Table \ref{tb:g_rho_psi}. The normalization factor '2' in Eq.~\eqref{eqn:skew_RES} arises naturally from the multiplication with the cdf of a RES distribution, because every cdf of a RES distribution has a mean value of $1/2$. This result can be easily verified by taking the symmetry of the RES distributions into account. Hence, the same result is obtained by evaluating  $F(0) = 1/2$. To simplify the notation, the argument of $\kappa\left(\underline{t}\right)$ is dropped and the following scalar variables are introduced:
\begin{equation}
\eta = \frac{\blambda^{\top} \bS^{-1}\left(\bx - \bxi\right)}{1 + \blambda^{\top}\bS^{-1}\blambda}, \qquad \tau^{2} = \frac{1}{1 + \blambda^{\top}\bS^{-1}\blambda}
\label{eqn:eta}
\end{equation}
Figure~\ref{fig:pdf_gaus_lambda} illustrates the flexibility in modeling skewed data by plotting the univariate skew-Gaussian distribution for different values of the skewness parameter $\lambda$. In the multivariate case, the skewness vector $\blambda$ allows for different amounts of skewness in every dimension. Figure~\ref{fig:pdf_gaus_lambda_xi} shows the influence of the centroid $\xi$, which for the univariate case simply shifts the distribution to the left or right. In the multivariate case, $\bxi$ shifts independently the distribution in each dimension. While the figure illustrates these parameters for the skew-Gaussian distribution, their meaning is identical for other RESK distributions.

We next briefly revisit some known special cases of RESK distributions before introducing the skew-Huber distribution. All required functions are summarized in Tables \ref{tb:g_rho_psi} and \ref{tb:eta_PSI}.

\subsection{Examples of Known Existing RESK Distributions}

\subsubsection{RES Distributions}
The RESK distribution has the property that, for $\blambda = \boldsymbol{0}$, it is equal to a RES distribution as defined in Eq.~\eqref{eqn:res}, because $\eta = 0$ , $\tau = 1$ and $\kappa\left(\underline{t}\right) = 0$, which leads to $F(0) = 1/2$. Also, one can find that $\bxi = \bmu$, hence the RESK distribution is a generalization of the RES distribution. This can be also observed in Figure~\ref{fig:pdf_gaus_lambda}. By comparing the black dashed curve with the red curve, it becomes clear that, for $\lambda = 0$, the skew-Gaussian is equal to the Gaussian distribution.

\subsubsection{Skew-Gaussian Distribution}
The pdf is defined by
\begin{equation}
\phi_{\text{s}}(\bx | \bxi, \bS, \blambda) = 2 \phi(\bx | \bxi, \bOmega) \Phi\left(\frac{\eta}{\tau}\right)
\label{eqn:skew_gaussian}
\end{equation}
with $\phi(\bx | \bmu, \bSigma)$ being the multivariate pdf and $\Phi(x)$ being the univariate cdf of the Gaussian distribution. 
\subsubsection{Skew-t Distribution}
The pdf is defined by
\begin{equation}
t_{\text{s}}(\bx | \bxi, \bS, \blambda, \nu) = 2 t(\bx | \bxi, \bOmega, \nu) T_{\nu + r}\left(\kappa\left(\underline{t}\right)\right),
\end{equation}
where $t(\bx | \bmu, \bPsi, \nu)$ is the multivariate pdf and $T_{\nu}(x)$ is the univariate cdf of a t distribution with associated degree of freedom parameter~$\nu$. During our work with the skew-t distribution we came across some inconsistencies regarding the definition of $\underline{t}$ and $\frac{\eta}{\tau}$ of the skew-t distribution between \cite{Vrbik.2014} and \cite{Pyne.2009b}. These inconsistencies are corrected here and the correct expressions are given in Eqs.~\eqref{eqn:skew_maha} and \eqref{eqn:eta}.

\subsection{Proposed Skew-Huber Distribution}
\label{sec:skew_huber}
Based on the definition in Eq.~\eqref{eqn:skew_RES}, the skew-Huber distribution can now be introduced as
\begin{equation}
h_{\text{s}}(\bx | \bxi, \bS, \blambda, c) = 2 h(\bx | \bxi, \bOmega, c) H_{c}\left(\kappa\left(\underline{t}\right)\right)
\end{equation}
with the multivariate Huber pdf from Eq. \eqref{eqn:huber_pdf} and the univariate Huber cdf from Eq.~\eqref{eqn:huber_cdf}, which are derived in the following.
\subsubsection{Multivariate pdf of Huber Distribution}
Huber's M-estimator can be seen as the ML estimator for a specific RES distribution (the Huber distribution) \cite[p.~115]{Zoubir.2018}, \cite[p.~8]{Ollila.2016b}. It is defined by
\begin{align}
h(t; r, c) =& \exp \left(-\frac{1}{2} \rho_{\text{H}}(t;c)\right)
\end{align}
with some tuning constant $c \geq 0 $ and
\begin{equation}
\rho_{\text{H}}(t;c) = 
\begin{dcases}
\frac{t}{b} &, t \leq c^{2}\\
\frac{c^{2}}{b}\left(\ln\left(\frac{t}{c^{2}}\right)+1\right) &, t > c^{2}
\end{dcases}.
\end{equation}
To obtain Fisher consistency 
\begin{equation}
b = F_{\chi_{r+2}^{2}}\left(c^{2}\right) + \frac{c^{2}}{r}\left(1 - F_{\chi_{r}^{2}}\left(c^{2}\right)\right),
\end{equation}
where $F_{\chi_{r}^{2}}\left(\cdot\right)$ is the Chi-square cumulative distribution function with degree of freedom $r$. For $h(t; r, c)$ to be a valid pdf the normalization factor, according to \cite{Sahu.2003}, has to be calculated as
\begin{align}
\int_{0}^{\infty}& u^{r/2-1} h(u; r, c)  \text{d}u \notag \\*
=& \int_{0}^{c^{2}} u^{r/2-1} e^{-\frac{u}{2b}} \text{d}u + \int_{c^{2}}^{\infty} u^{r/2-1} \left(\frac{e u}{c^{2}} \right)^{-\frac{c^{2}}{2b}} \text{d}u \notag \\
=& (2b)^{r/2}\left(\Gamma\left(\frac{r}{2}\right) - \Gamma\left(\frac{r}{2}, \frac{c^{2}}{2b}\right)\right) + \frac{2 b c^{r} \exp \left(-\frac{c^{2}}{2b} \right)}{c^2 - br},
\end{align}
with the gamma function $\Gamma(\cdot)$, the upper incomplete gamma function $\Gamma(\cdot, \cdot)$ and Euler's number $e$. The density generator, see Eq.~\eqref{eqn:res}, of a Huber distribution now follows as
\begin{align}
g(t) =& 
\begin{dcases}
A_{\text{H}} \exp \left(-\frac{t}{2b}\right) &, t \leq c^{2}\\
A_{\text{H}} \left(\frac{e t}{c^{2}} \right)^{-\frac{c^{2}}{2b}} &, t > c^{2}
\end{dcases},
\end{align}
with
\begin{equation}
\resizebox{\columnwidth}{!}{$
	A_{\text{H}} = \frac{\Gamma\left(\frac{r}{2}\right)}{\pi^{r/2}} \left((2b)^{\frac{r}{2}}\left(\Gamma\left(\frac{r}{2}\right) - \Gamma\left(\frac{r}{2}, \frac{c^{2}}{2b}\right)\right)+\frac{2 b c^{r} e^{-\frac{c^{2}}{2b} }}{c^2 - br}\right)^{-1}$}.
\end{equation}
Then, the multivariate Huber pdf is obtained as
\begin{align}
h(\bx | \bmu, \bS, c) =& 
\begin{dcases}
\left| \bS \right|^{-\frac{1}{2}} A_{\text{H}} \exp \left(-\frac{t}{2b}\right) &, t \leq c^{2}\\
\left| \bS \right|^{-\frac{1}{2}} A_{\text{H}} \left(\frac{e t}{c^{2}} \right)^{-\frac{c^{2}}{2b}} &, t > c^{2}
\end{dcases}
\label{eqn:huber_pdf}
\end{align}

\subsubsection{Univariate cdf of Huber Distribution}
We start with the univariate pdf
\begin{align}
h(x | \mu, s, c) =& 
\begin{dcases}
\frac{A_{\text{H}}}{s} e^{-\frac{(x - \mu)^{2}}{2b s^{2}}} &, \left|\frac{x-\mu}{s}\right|\leq c\\
\frac{A_{\text{H}}}{s} \left(\frac{e (x-\mu)^{2}}{c^{2}s^{2}} \right)^{-\frac{c^{2}}{2b}} &, \left|\frac{x-\mu}{s}\right| > c
\end{dcases}
\end{align}
and integrate it. This leads to
\begin{align}
H&_{c}\left(z\right)= \notag \\* 
&\begin{dcases}
\frac{A_{\text{H}} b z}{b - c^{2}} \left(\frac{e z^{2}}{c^{2}}\right)^{-\frac{c^{2}}{2b}} &, z < -c\\
\begin{aligned}
&\frac{- A_{\text{H}} b c}{b - c^{2}} e^{-\frac{c^{2}}{2b}} 
\\& \quad + A_{\text{H}} \sqrt{\frac{\pi b }{2}} \left(\erf\left(\frac{z}{\sqrt{2 b}}\right) + \erf\left(\frac{c}{\sqrt{2 b}}\right)\right)
\end{aligned}&,  |z| \leq c\\
\begin{aligned}
&\frac{- A_{\text{H}} b c}{b - c^{2}} e^{-\frac{c^{2}}{2b}} + A_{\text{H}} \sqrt{2 \pi b} \erf\left(\frac{c}{\sqrt{2 b}}\right) 
\\& \quad + \frac{A_{\text{H}} b}{b - c^{2}} \left(\frac{e}{c^{2}} \right)^{-\frac{c^{2}}{2b}} \left(z \left(z^{2}\right)^{-\frac{c^{2}}{2b}} - c^{1-\frac{c^{2}}{b}}\right)
\end{aligned}&,  z > c
\end{dcases}
\label{eqn:huber_cdf}
\end{align}
with $z = \frac{x-\mu}{s}$ and the integrals
\begin{align}
\int_{-\infty}^{z} & A_{\text{H}}  \left(\frac{e u^{2}}{c^{2}} \right)^{-\frac{c^{2}}{2b}} du \notag \\ 
=& A_{\text{H}} \left(\frac{e}{c^{2}} \right)^{-\frac{c^{2}}{2b}} \frac{b}{b - c^{2}} \left[u\left(u^{2} \right)^{-\frac{c^{2}}{2b}}\right]_{-\infty}^{z} \notag \\
=& A_{\text{H}} \left(\frac{e}{c^{2}} \right)^{-\frac{c^{2}}{2b}} \frac{b}{b - c^{2}} \left(z\left(z^{2} \right)^{-\frac{c^{2}}{2b}} - \lim\limits_{u \rightarrow - \infty} u\left(u^{2} \right)^{-\frac{c^{2}}{2b}}\right) \notag \\
=& A_{\text{H}} \left(\frac{e}{c^{2}} \right)^{-\frac{c^{2}}{2b}} \frac{b}{b - c^{2}} z \left(z^{2} \right)^{-\frac{c^{2}}{2b}}
\label{eqn:hub_int}
\end{align}
with the constraint $\frac{c^{2}}{2b} > 1$ and
\begin{align}
\int_{-c}^{z} A_{\text{H}} & \exp \left(-\frac{u^{2}}{2b}\right) du \notag \\
=&  A_{\text{H}} \sqrt{\frac{\pi b }{2}} \left(\erf\left(\frac{z}{\sqrt{2 b}}\right) + \erf\left(\frac{c}{\sqrt{2 b}}\right)\right).
\end{align}
When deriving estimators based on the Huber distribution, \cite[p.~116]{Zoubir.2018} suggest to choose $c^2$ as the $q_H$th upper quantile of a $\chi_{r}^{2}$ distribution
\begin{equation}
c^{2} = F_{\chi_{r}^{2}}^{-1}\left(q_{H}\right), \quad 0 < q_{H} < 1.
\label{eqn:qH}
\end{equation}
The value of $q_{H}$ is chosen to trade-off efficiency under a Gaussian model and robustness against outliers. We can proceed similarly for the skew-Huber, but from $\frac{c^{2}}{2b} > 1$ in Eq.~\eqref{eqn:hub_int}, we have the constraint that $q_{H} > 0.703$. In Figure~\ref{fig:pdf_huber_q}, the skew-Huber distribution is plotted over different values of $q_{H}$ from Eq. \eqref{eqn:qH}. For lower values of $q_{H}$, probability mass is shifted to the tails of the distribution. As $q_{H} \rightarrow 1$ and $c~\rightarrow~\infty$ the skew-Huber becomes equal to the skew-Gaussian distribution. Additionally, in Figure~\ref{fig:psi_gaus_t_huber} the influence of the weight function $\psi(\underline{t})$ is shown. Similarly, as $q_{H} \rightarrow 1$ and $c~\rightarrow~\infty$ the weight function becomes equal to the Gaussian weight function.
\begin{table*}[t]
	\renewcommand{\arraystretch}{1.3}
	\caption{Overview of $g(t_{n})$, $\rho(t_{n})$ and $\psi(t_{n})$ functions}
	\label{tb:g_rho_psi}
	\centering
	{\tabulinesep=1.7mm
		\begin{tabu}{cccc}
			& $g(t_{n})$ & $\rho(t_{n})$ & $\psi(t_{n})$ \\\hline
			Gaussian
			& $(2 \pi)^{-\frac{r}{2}} \exp \left(-\frac{1}{2} t_{n}\right)$ 
			& $\frac{1}{2} t_{n} + \frac{r}{2} \ln\left(2 \pi\right)$ 
			& $\frac{1}{2}$ \\
			t
			& $\frac{\Gamma\left((\nu+r)/2\right)}{\Gamma\left(\nu/2 \right)(\pi \nu)^{r/2}} \left(1 + \frac{t_{n}}{\nu}\right)^{-(\nu + r)/2}$ 
			& $-\ln\left(\frac{\Gamma\left((\nu+r)/2\right)}{\Gamma\left(\nu/2 \right)(\pi \nu)^{r/2}}\right) + \frac{\nu + r}{2} \ln\left(1 + \frac{t_{n}}{\nu}\right)$ 
			& $\frac{1}{2} \cdot\frac{\nu + r}{\nu + t_{n}}$ \\
			Huber
			& $\begin{dcases}
			A_{\text{H}} \exp \left(-\frac{t_{n}}{2b}\right) &, t_{n} \leq c^{2}\\
			A_{\text{H}} \left(\frac{t_{n}}{c^{2}} \right)^{-\frac{c^{2}}{2b}} \exp \left(-\frac{c^{2}}{2b} \right) &, t_{n} > c^{2}
			\end{dcases}$ 
			& $	\begin{dcases}
			-\ln\left(A_{\text{H}}\right) + \frac{t_{n}}{2b} &, t_{n} \leq c^{2}\\
			-\ln\left(A_{\text{H}}\right) + \frac{c^{2}}{2b} \left(\ln\left(\frac{t_{n}} {c^{2}}\right)+1\right) &, t_{n} > c^{2}
			\end{dcases}$ 
			& $	\begin{dcases}
			\frac{1}{2b} &, t_{n} \leq c^{2}\\
			\frac{c^{2}}{2bt_{n}} &, t_{n} > c^{2}
			\end{dcases}$ \\ \hline
	\end{tabu}}
\end{table*}
\begin{table}[t]
	\caption{Overview of $\eta(t_{n})$ and $\Psi(x)$ functions}
	\label{tb:eta_PSI}
	\centering
	\begin{threeparttable}
		{\renewcommand{\arraystretch}{2}
			\begin{tabular}{ccc}
				& $\eta(t_{n})$ & $\Psi(x)$ \\\hline
				Gaussian
				& $0$ 
				& $ -\frac{\phi(x|0, 1)}{\Phi(x)}$ \tnote{*} \\
				t
				& $- \frac{1}{2} \cdot \frac{\nu + r}{(\nu + t_{n})^{2}}$
				& $ -\frac{t(x|0, 1, \nu + r)}{T_{\nu + r}(x)}$ \\[10pt]
				Huber
				& $	\begin{dcases}
				0 &, t_{n} \leq c^{2}\\
				-\frac{c^{2}}{2bt_{n}^{2}} &, t_{n} > c^{2}
				\end{dcases}$ 
				& $ -\frac{h(x|0, 1, c)}{H_{c}(x)}$\\[15pt]
				\hline
		\end{tabular}}
	\begin{tablenotes}\footnotesize
		\item[*] The calculation of $\Psi(x)$ can be numerically unstable, due to the fact that $\Phi(x)$ tends to zero for $x < -37$. An approximation for these values can be found in \cite[29]{Vrbik.2014}.
	\end{tablenotes}
	\end{threeparttable}
\end{table}

\section{Proposed Expectation Maximization Algorithm for a Mixture of RESK Distributions}
\label{ap:EM_skew}
In this section, we derive an EM algorithm to estimate the RESK mixture model parameters and the cluster memberships of the data vectors $\xn$ by alternating the E- and M-steps until convergence. The proposed approach is summarized in Algorithm~\ref{alg:em_skew}, where we provide a unified EM clustering framework that is beneficial for skewed and heavy-tailed data sets. The user simply chooses a RESK density generator $g(t)$ and determines the associated weight function $\psi(t)$. We have implemented some specific variants of the EM algorithm, for example the skew-Huber, based on the values from Tables~\ref{tb:g_rho_psi} and \ref{tb:eta_PSI}. However, the framework is not limited to these cases, and therefore it is very flexible in modeling different data distributions.

The parameter vector, for each cluster $m=1,\dots, l$, is defined as $\btheta_{m} = \left[\bxi_{m}^{\top}, \blambda_{m}^{\top}, \vechop(\bS_{m})^{\top}\right]^{\top} \in \mathbb{R}^{q \times 1}$, where $l$ is the number of clusters and $q = \frac{r}{2}(r+5)$ is the number of parameters per cluster. Because $\bS_{m}$ is symmetric, it has only $\frac{r}{2}(r+1)$ unique elements. Therefore, $\vechop(\bS_{m})$ has to be used in the estimation step \mbox{{\cite[p.~367]{Abadir.2005}}}. Accordingly, we derive an EM algorithm for RESK distributions in the following.

Based on Equation~\eqref{eqn:skew_RES}, the mixture of $l$ RESK distributions is formulated as
\begin{align}
	p(\bx | \bPhi_{l}) = \sum_{m=1}^{l} \gamma_{m} f_{\text{s}}(\bx | \btheta_{m}) 
	\label{eqn:mixture_skew}
\end{align}
and the incomplete data log-likelihood follows as
\begin{align}
\ln\left(\mathcal{L}(\bPhi_{l}|\mathcal{X} )\right) =& \ln\left( p(\mathcal{X} | \bPhi_{l})\right)\notag\\
=& \sum_{n = 1}^{N} \ln \left(\sum_{m=1}^{l} \gamma_{m} f_{\text{s}}(\bx | \btheta_{m}) \right)
\label{eqn:emml_skew}
\end{align}
with $\gamma_{m}$ being the mixing coefficient and  $\bPhi_{l} = [\bgamma_{l}, \bTheta_{l}^{\top}]$ with $ \bgamma_{l} = [\gamma_{1},\dots, \gamma_{l}]^{\top}$ and $\bTheta_{l} = [\btheta_{1},\dots,\btheta_{l}] \in \mathbb{R}^{q \times l}$.

Following the derivation of the EM algorithm from \cite{Bishop.2009}, a latent binary variable $\bz = [z_{1},\dots, z_{l}]^{\top}$ is introduced, which has a single component equal to 1 and all other components equal to 0. The conditional distribution of $\bx$ given $\bz$, becomes
\begin{equation}
	p(\bx | \bz, \bTheta_{l}) = \prod_{m=1}^{l} f_{\text{s}}(\bx | \btheta_{m})^{z_{m}},
\end{equation}
where the prior of the latent variables is given by
\begin{equation}
	p(\bz | \bgamma_{l}) = \prod_{m=1}^{l} \gamma_{m}^{z_{m}}
\end{equation}
and the joint distribution equals
\begin{equation}
	p(\bx, \bz | \bPhi_{l}) = \prod_{m=1}^{l} \left(\gamma_{m} f_{\text{s}}(\bx | \btheta_{m})\right)^{z_{m}}.
\end{equation}
Now, the complete data log-likelihood, with $\mathcal{X}$ and $\mathcal{Z} = \{\bz_{1},\dots,\bz_{N}\}$, is formulated as
\begin{align}
	\ln(p(\mathcal{X}, \mathcal{Z} | &\bPhi_{l})) \notag \\
	=& \ln\left(\prod_{n=1}^{N} \prod_{m=1}^{l}\left(\gamma_{m} f_{\text{s}}(\bx_{n} | \btheta_{m})\right)^{z_{nm}}\right) \notag \\
	=& \sum_{n=1}^{N} \sum_{m=1}^{l} z_{nm} \left(\ln(\gamma_{m}) + \ln(f_{\text{s}}(\bx_{n} | \btheta_{m}))\right),
\end{align}
where $z_{nm}$ denotes the $m$th element of $\bz_{n}$. For the E-step the expected value of the complete data log-likelihood with respect to the latent variables is calculated as
\begin{align}
	E_{\mathcal{Z}}[\ln(p(&\mathcal{X}, \mathcal{Z} | \bPhi_{l}))] = Q(\bPhi_{l}) \notag \\
	&= \sum_{n=1}^{N} \sum_{m=1}^{l} v_{nm} \left(\ln(\gamma_{m}) + \ln(f_{\text{s}}(\bx_{n} | \btheta_{m}))\right)
	\label{eqn:complete}
\end{align}
with
\begin{align}
	v_{nm} = E[z_{nm}] =& \sum_{z_{nm}} z_{nm} p(z_{nm}| \bx_{n}, \gamma_{m}, \btheta_{m}) \notag \\
	=& \sum_{z_{nm}}  \frac{z_{nm} p(z_{nm} | \gamma_{m}) p(\bx_{n} | z_{nm}, \btheta_{m})}{p(\bx_{n}|\gamma_{m}, \btheta_{m})}\notag \\
	=&   \frac{\sum\limits_{z_{nm}} z_{nm}  (\gamma_{m} f_{\text{s}}(\bx_{n} | \btheta_{m}))^{z_{nm}}} {\sum\limits_{j=1}^{l} \gamma_{j} f_{\text{s}}(\bx_{n} | \btheta_{j}) }\notag \\
	=&   \frac{\gamma_{m} f_{\text{s}}(\bx_{n} | \btheta_{m})} {\sum\limits_{j=1}^{l} \gamma_{j} f_{\text{s}}(\bx_{n} | \btheta_{j}) }.
	\label{eqn:vskew}
\end{align}
The calculation of the ML estimates when using the EM algorithm requires fixed values for the variable $v_{nm}$ and additionally for $e_{0,nm}$, $e_{1,nm}$ and $e_{2,nm}$ from Eqs. \eqref{eqn:e0}, \eqref{eqn:e1} and \eqref{eqn:e2}. In the E-step of the EM algorithm $v_{nm}$ is fixed with respect to $\btheta_{m}$ by calculating the expected value of $z_{nm}$. A latent variable approach would be also favorable for $e_{0,nm}$, $e_{1,nm}$ and $e_{2,nm}$, but it turns out, that for general RESK distributions this is not feasible. Hence, $e_{0,nm}$, $e_{1,nm}$ and $e_{2,nm}$ are approximated by the estimated values $\bthetahat_{m}$ of the previous iteration of the EM algorithm into Eqs. \eqref{eqn:e0}, \eqref{eqn:e1} and \eqref{eqn:e2} which yields the approximations $\tilde{e}_{0,nm}$, $\tilde{e}_{1,nm}$ and $\tilde{e}_{2,nm}$. If the approximations hold exactly, which is the case e.g. for the skew-Gaussian distribution, the ML estimate is obtained. Otherwise an approximate ML estimate is obtained. In the context of M-estimation, where the primary aim is trading off near-optimality at a nominal distribution for robustness over a family of distributions \cite{Zoubir.2018, Huber.2011}, such an approximation is well motivated and of high practical value. Based on the approximations, the expected value of the complete data log-likelihood, is maximized, which results in the following estimates for the M-step:
\begin{align}
	\bxihat_{m} = \frac{\sum_{n = 1}^{N} v_{nm} \left(\tilde{e}_{0,nm} \bx_{n}  - \tilde{e}_{1,nm}  \blambda_{m}\right)}{\sum_{n = 1}^{N} v_{nm} \tilde{e}_{0,nm} },
\end{align}
\begin{align}
	\blambdahat_{m} = \frac{\sum_{n = 1}^{N} v_{nm} \tilde{e}_{1,nm} \bxt_{n}}{\sum_{n = 1}^{N} v_{nm} \tilde{e}_{2,nm} }
\end{align}
with $\bxt_{n} = \bx_{n} - \bxi_{m}$,
\begin{align}
	\begin{split}
		& \bShat_{m} = \Biggl[\sum_{n = 1}^{N} v_{nm} \Bigl(\tilde{e}_{0,nm} \bxt_{n}\bxt_{n}^{\top}  - \tilde{e}_{1,nm}\bxt_{n} \blambda_{m}^{\top} 
		\\&- \tilde{e}_{1,nm}\blambda_{m}\bxt_{n}^{\top} + \tilde{e}_{2,nm} \blambda_{m} \blambda_{m}^{\top} \Bigr)\Biggr]\biggm/ \sum_{n = 1}^{N} v_{nm},
	\end{split}
\end{align}
and
\begin{align}
	\hat{\gamma}_{m} =& \frac{1}{N} \sum_{n = 1}^{N} v_{nm}.
\end{align}
A detailed derivation of the M-step is given in Appendix~\ref{sec:app}.
\begin{figure}[!ht]
	\removelatexerror
	\begin{algorithm}[H]
		\KwIn{$\mathcal{X}$, $i_{\max}$, $l$, $g(t)$, $\psi(t)$}
		\KwOut{$\bxihat_{m}$, $\bShat_{m}$, $\blambdahat_{m}$, $\hat{\gamma}_{m}$}
		\For{$m = 1,\dots, l$}
		{
			Initialize $\bxihat_{m}^{(0)}$, e.g. with K-means or K-medoids\;
			Initialize $\blambdahat_{m}^{(0)} = \boldsymbol{1}$\;
			\[
			\bShat_{m}^{(0)} = \frac{1}{N_m}\sum_{\xinX} \left(\xn - \bxihat_{m}^{(0)}\right) \left(\xn - \bxihat_{m}^{(0)}\right)^{\top}
			\]
			\[
			\hat{\gamma}_{m}^{(0)} = \frac{N_{m}}{N}
			\]
		}
		\For{$i = 1,\dots, i_{\max}$}
		{%
			E-step:\;
			\For{$m = 1,\dots, l$}
			{
				\For{$n = 1,\dots, N$}
				{
					Calculate $\hat{v}_{nm}^{(i)}$, $\tilde{e}_{0,nm}^{(i)}$, $\tilde{e}_{1,nm}^{(i)}$ and $\tilde{e}_{2,nm}^{(i)}$ with Eqs. \eqref{eqn:vskew}, \eqref{eqn:e0}, \eqref{eqn:e1} and \eqref{eqn:e2} using the estimates from the previous iteration
			}}
			M-Step:\;
			\For{$m = 1,\dots, l$}
			{
				\[
				\bxihat_{m}^{(i)} = \frac{\sum_{n = 1}^{N} \hat{v}_{nm}^{(i)} \left(\tilde{e}_{0,nm}^{(i)} \bx_{n} - \tilde{e}_{1,nm}^{(i)}  \blambdahat_{m}^{(i-1)}\right)}{\sum_{n = 1}^{N} \hat{v}_{nm}^{(i)} \tilde{e}_{0,nm}^{(i)} }
				\]
				\[
				\blambdahat_{m}^{(i)} = \frac{\sum_{n = 1}^{N} \hat{v}_{nm}^{(i)} \tilde{e}_{1,nm}^{(i)} \left(\xn - \bxihat_{m}^{(i)}\right)}{\sum_{n = 1}^{N} \hat{v}_{nm}^{(i)} \tilde{e}_{2,nm}^{(i)} } 
				\]				
				\[
				\begin{split}
				\bShat&_{m}^{(i)} = \Biggl[\sum_{n = 1}^{N} \hat{v}_{nm}^{(i)} \Bigl[\tilde{e}_{0,nm}^{(i)} \bxhat_{n}^{(i)}\left(\bxhat_{n}^{(i)}\right)^{\top}  
				\\&- \tilde{e}_{1,nm}^{(i)}\bxhat_{n}^{(i)} \left(\blambdahat_{m}^{(i)}\right)^{\top} - \tilde{e}_{1,nm}^{(i)}\blambdahat_{m}^{(i)}\left(\bxhat_{n}^{(i)}\right)^{\top} 
				\\&+ \tilde{e}_{2,nm}^{(i)} \blambdahat_{m}^{(i)} \left(\blambdahat_{m}^{(i)}\right)^{\top} \Bigr]\Biggr]\biggm/ \sum_{n = 1}^{N} \hat{v}_{nm}^{(i)}
				\end{split}
				\]
				\[
				\text{with } \bxhat_{n}^{(i)} = \bx_{n} - \bxihat_{m}^{(i)}
				\]
				\[
				\hat{\gamma}_{m}^{(i)} = \frac{1}{N} \sum_{n = 1}^{N} \hat{v}_{nm}^{(i)}
				\]
			}
			Calculate log-likelihood:
			\[
			\begin{split}
			\ln(\mathcal{L}(\bPhihat_{l}^{(i)}|\mathcal{X})) = \sum_{n = 1}^{N} \ln \Bigl(&\sum_{m=1}^{l} \hat{\gamma}_{m}^{(i)} 2 \left| \bOmegahat_{m}^{(i)} \right|^{-\frac{1}{2}} 
			\\&\times g\left( \underline{\that}_{nm}^{(i)}\right) F\left(\hat{\kappa}_{nm}^{(i)}\right) \Bigr)
			\end{split}
			\]
			\If{$\left|\ln\left(\mathcal{L}\left(\bPhihat_{l}^{(i)}|\mathcal{X} \right)\right)-\ln\left(\mathcal{L}\left(\bPhihat_{l}^{(i-1)}|\mathcal{X} \right)\right)\right| < \delta$}
			{
				\;
				break loop
			}
		}
		\caption{EM algorithm for RESK distribution}
		\label{alg:em_skew}
	\end{algorithm}
\end{figure}

\section{Cluster Enumeration}
\label{sec:clu_enum}
This section gives a brief overview of the Schwarz BIC \cite{Schwarz.1978, Cavanaugh.1999}, which we use to estimate the number of clusters. There exist alternative criteria, e.g. \cite{Teklehaymanot.2018, Teklehaymanot.2018b, Schroth.2021}, but in this paper the focus does not lie on cluster enumeration. Hence, only the log-likelihood function is adapted to RESK distributions and the Schwarz BIC is applied, i.e.
\begin{align}
\text{BIC}_{\text{o}}(M_{l})=& \sum_{m=1}^{l} \ln\left(\mathcal{L} \left(\bthetahat_{m}|\mathcal{X}_{m}\right)\right) - \frac{ql}{2} \ln\left(N\right)
\label{eqn:bic_schwarz}
\end{align}
with $M_{l}$ being a candidate model with $l \in \{L_{\min},\dots,L_{\max}\}$ clusters. The true number of clusters $K$ is assumed to lie within $L_{\min} \leq K \leq L_{\max} $. For the RES case, the log-likelihood function is
\begin{align}
\begin{split}
\ln\left(\mathcal{L}(\bthetahat_{m}|\mathcal{X}_{m} )\right) =& -\sum_{\xinX} \rho(\that_{nm}) + N_{m} \ln \left(\frac{N_{m}}{N}\right) 
\\&- \frac{N_{m}}{2} \ln\left(\left| \bShat_{m}\right|\right).
\end{split}
\label{eqn:loglikelihood}
\end{align}
For the RESK case, the log-likelihood function is derived as
\begin{align}
\ln(\mathcal{L}&(\bthetahat_{m}|\mathcal{X}_{m} )) \notag \\
\begin{split}
=&-\sum_{\xinX} \rho(\underline{\that}_{nm}) + N_{m} \ln \left(\frac{N_{m}}{N}\right) + N_{m} \ln \left(2\right)
\\&- \frac{N_{m}}{2} \ln\left(\left| \bOmegahat_{m}\right|\right)  + \sum_{\xinX} \ln\left( F\left(\hat{\kappa}_{nm}\right)\right).
\end{split}
\label{eqn:loglikelihood_skew}
\end{align}
The number of clusters is estimated by evaluating
\begin{equation}
	\hat{K} = \argmax_{l = L_{\min},\dots,L_{\max}}  \text{BIC}_{\text{o}}(M_{l}).
	\label{eqn:K_hat_ares}
\end{equation}
\section{Experimental Results}
\label{sec:sim}
This section reports on our numerical experiments to evaluate the proposed EM algorithms based on RESK distributions in terms of (i) their parameter estimation and clustering performance for skewed and heavy-tailed mixture models; (ii) their robustness against outliers; (iii) their convergence speed in terms of the number of required iterations; and (iv) their ability to model real-world data. We compare the newly proposed skew-Huber estimator to other known estimators, such as the skew-Gaussian, skew-t \cite{Vrbik.2014}, and the classic Gaussian, t- and Huber estimators. An implementation of the proposed EM algorithm and the conducted simulations is available at: \url{https://github.com/schrchr/Real-Elliptically-Skewed-Distributions}.

\subsection{Dataset 1}
\begin{figure}[t]
	\centering
	\subfloat[Simulated data with \mbox{$\epsilon = 2\%$} replacement outliers.] {\resizebox{0.48\columnwidth}{!}{\begin{tikzpicture}
	\begin{axis}[
	/pgf/number format/.cd,
	fixed,
	1000 sep={},
	width = \figurewidth,
	height = \figureheight,
	xmin = -15,
	xmax = 45,
	ymin = -20,
	ymax = 30,
	xlabel= Feature 1,
	ylabel= Feature 2,
	legend entries={{cluster 1}\\
		{cluster 2}\\
		{cluster 3}\\
		{outliers}\\},
	legend pos=north west,
	legend cell align={left},
	]
	\addplot[scatter/classes={1={matlabblue},2={matlaborange},3={matlabyellow}, 4={black, mark = o}}, scatter, only marks, scatter src=explicit symbolic] table[x=x,y=y,meta=label] {figures/data_31_skew_50_eps_002.csv};   
	\end{axis}
\end{tikzpicture}}%
		\label{fig:data_31_skew_002}}
	\hfil
	\subfloat[First three features of the wine data.] {\resizebox{0.48\columnwidth}{!}{\begin{tikzpicture}
	\begin{axis}[
	/pgf/number format/.cd,
	fixed,
	1000 sep={},
	width = \figurewidth,
	height = \figureheight,
	xlabel= volatile acidity,
	ylabel= residual sugar,
	zlabel= chlorides,
	legend entries={{white wine}\\
		{red wine}\\},
	legend pos=north east,
	legend cell align={left},
	view={130}{20}, 
	]
	\addplot3[scatter/classes={1={matlabblue,mark size=3pt},2={matlaborange,mark size=3pt},3={matlabyellow,mark size=3pt}, 4={matlabpurple,mark size=3pt}}, scatter, only marks, scatter src=explicit symbolic] table[x=x,y=y,z=z,meta=label] {figures/data_red_white_wine.csv};   
	\end{axis}
\end{tikzpicture}}%
		\label{fig:data_red_white_wine}}
	\caption{Exemplary realizations of the simulated data set and the wine data set.}
\end{figure}
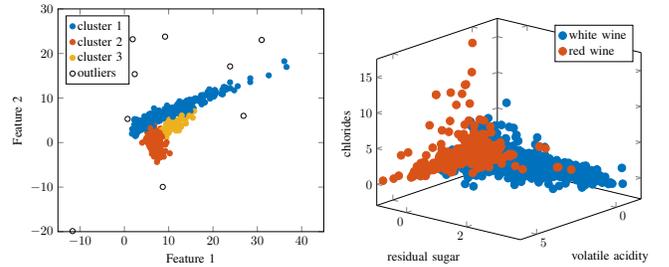

The first data set is comprised of $K=3$ skew-Gaussian distributed clusters. It is generated with the algorithm from \cite[23]{Vrbik.2014}, and $\bx_{k} \sim \mathcal{N}_{s}\left(\bxi_{k}, \bSigma_{k}, \blambda_{k}\right)$, $k = 1,2,3$. The cluster centroids are chosen as $\bxi_{1} = [2, 3.5]^{\top}$, $\bxi_{2} = [6, 2]^{\top}$ and $\bxi_{3} = [10, 3]^{\top}$, the skewness factors are $\blambda_{1} = [10, 4]^{\top}$, $\blambda_{2} = [1, -2]^{\top}$ and $\blambda_{3} = [2, 1]^{\top}$ and the covariance matrices are given by
\begin{equation}
\bSigma_{1} = \begin{bmatrix}
0.2 & 0.1\\
0.1 & 0.75\end{bmatrix}, 
\bSigma_{2} = \begin{bmatrix}
0.5 & 0.25\\
0.25 & 0.5\end{bmatrix},
\bSigma_{3} = \begin{bmatrix}
1 & 0.5\\
0.5 & 1\end{bmatrix}. \notag
\end{equation}
The number of data points per cluster is specified as $N_{1} = 5 \times N_{K}$, $N_{2} = 4 \times N_{K}$ and $N_{3} = 1 \times N_{K}$. An exemplary realization is given in Figure~\ref{fig:data_31_skew_002}. 

\subsection{Dataset 2}
The second data set comprised of $K=2$ skew-t distributed clusters is generated with the algorithm from \cite[23]{Vrbik.2014}, and $\bx_{k} \sim \mathcal{T}_{s}\left(\bxi_{k}, \bS_{k}, \blambda_{k}, \nu_{k}\right)$, $k = 1,2,3$. The cluster centroids are chosen as $\bxi_{1} = [2, 3.5]^{\top}$ and $\bxi_{2} = [7, -2]^{\top}$, the skewness factors are $\blambda_{1} = [4, 3]^{\top}$ and $\blambda_{2} = [1, -2]^{\top}$, the degrees of freedom are set to $\nu_{1} = \nu_{2} = 3$ and the covariance matrices are given by
\begin{equation}
	\bS_{1} = \begin{bmatrix}
		0.2 & 0.1\\
		0.1 & 0.75\end{bmatrix}, 
	\bS_{2} = \begin{bmatrix}
		0.5 & 0.25\\
		0.25 & 0.5\end{bmatrix}. \notag
\end{equation}
The number of data points per cluster is specified as $N_{1} = N_{2} = N_{K}$.

\subsection{Outlier Generation}
To study robustness against outliers, an $\epsilon$-contamination model is used, where $\epsilon$ denotes the percentage of randomly replaced data points. To make the clustering problem well-defined, it is assumed that the outliers do not form a cluster. Therefore, a uniform distribution is chosen as the outlier generating distribution because it produces isolated data points that are placed with equal probability on a sufficiently large area compared to the clustered data. For Dataset 1 and Dataset 2, the outliers are uniformly distributed in the range of \mbox{[-15, 45]}, \mbox{[-20, 30]} in x, y-dimension.

\subsection{Choice of Tuning Parameters}
Based on \cite[121]{Zoubir.2018}, for Huber's and skew-Huber's estimator, a value of $q_{H} = 0.8$ is chosen for all simulations, which leads to $c = 1.282$. For all t and skew-t distributions, $\nu = 3$ is selected for the degree of freedom. Both are common choices in M-estimation to trade-off robustness against outliers and efficiency for Gaussian data. To evaluate the choice of $\nu = 3$, we also report on an oracle version, where the degree of freedom is selected from the set $\nu = \{1,2,3,4,5,7,9\}$ so that it provides the optimal performance.

\subsection{Performance Measures}

\subsubsection{Kullback-Leibler Divergence}
To evaluate the parameter estimation performance of the proposed method, we use the Kullback-Leibler (KL) divergence, which measures the difference between two pdfs \cite[p.~34]{MacKay.2011}. It is defined as
\begin{equation}
D_{KL}(p||q) = \sum_{i=1}^{N} p(x_{i}) \ln\left(\frac{p(x_{i})}{q(x_{i})}\right)
\end{equation}
with $p(x_{i})$ representing the pdf that generates the outlier-free data and $q(x_{i})$ representing the pdf estimated by the EM algorithm. Even if there are outliers in the data, the aim of robust methods is to provide information about $p(x_{i})$. Therefore, the KL divergence between $p(x_{i})$ and $q(x_{i})$ is computed.

\subsubsection{Confusion Matrix}
To evaluate the correctness in labeling the non-outlying data points, the confusion matrix for these points is reported.

\subsubsection{Sensitivity Curve}
The sensitivity curve (also known as empirical influence function \cite{Zoubir.2018}) shows the effect of a single outlier as a function of its position in the feature space. The concept of looking at single outliers is related to infinitesimal robustness, or qualitative robustness, as measured by the influence function. A robust method is expected to provide a nearly optimal performance and is not largely influenced by a single outlier, irrespective of the value that the outlier takes. In our experiments, the sensitivity curves are calculated as follows. One randomly chosen sample is replaced with an outlier from the range \mbox{[-15, 45]}, \mbox{[-20, 30]} in x, y-dimension. For each point of a grid of outlier values, the proposed EM algorithm is executed. The resulting estimated pdf is then compared with the true (outlier-free) generating pdf by calculating the KL divergence, which is averaged over 200 Monte Carlo iterations with $N_{K} = 50$.

\subsubsection{Breakdown Behavior}
Showing robustness against a single outlier is not sufficient in cluster analysis, where we may have positive a fraction of data points that do not follow the cluster structure of the remaining data points. This is several experiments with varying amounts of outliers are conducted. To analyze robustness against a fraction of outliers, data is generated from a mixture model with a nominal distribution (e.g. the Gaussian or the t) and a contaminating (outlier generating) distribution. The level of contamination is increased until all methods break down.

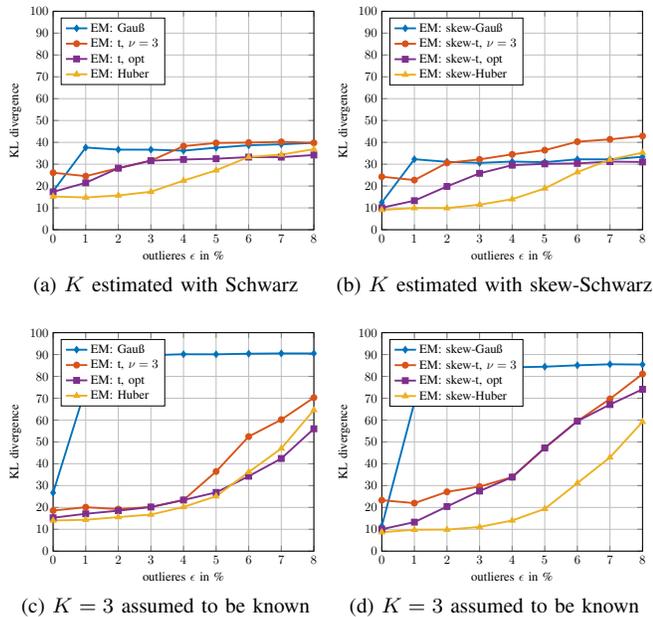
\begin{figure}[t]
	\centering
	\subfloat[$K$ estimated with Schwarz]
	{\resizebox{0.48\columnwidth}{!}{\begin{tikzpicture}
\begin{axis}[
	scaled ticks=false,
	tick label style={/pgf/number format/.cd},
	width = \figurewidth,
	height = \figureheight,
	xmin = 0,
	xmax = 0.08,
	ymax = 100,
	ymin = 0,
	ytick={0,10,20,30,40,50,60,70,80,90,100},
	yticklabels={$0$,$10$,$20$,$30$,$40$,$50$,$60$,$70$,$80$,$90$,$100$},
	xtick={0,0.01,0.02,0.03,0.04,0.05,0.06,0.07,0.08},
	xticklabels={$0$,$1$,$2$,$3$,$4$,$5$,$6$,$7$,$8$},
	grid,
	xlabel= outlieres $\epsilon$ in \%,
	ylabel= KL divergence,
	legend entries={{EM: Gauß}\\
		{EM: t, $\nu = 3$}\\
		{EM: t, opt}\\
		{EM: Huber}\\},
	legend pos=north west,
	legend cell align={left},
	cycle list name=matlabcolor
	]
	\addplot+[thick,matlabblue,line width=1.5pt] table[x index=0,y index=1,col sep=tab] {figures/t-opt/KL_BIC-Schwarz_MC_500_Nk_50.csv};
	\addplot+[thick,matlaborange,line width=1.5pt] table[x index=0,y index=5,col sep=tab] {figures/t-opt/KL_BIC-Schwarz_MC_500_Nk_50.csv};
	\addplot+[thick,matlabpurple,line width=1.5pt] table[x index=0,y index=3,col sep=tab] {figures/t-opt/KL_t_optBIC-Schwarz_MC_500_Nk_50.csv};
	\addplot+[thick,matlabyellow,line width=1.5pt] table[x index=0,y index=2,col sep=tab] {figures/t-opt/KL_BIC-Schwarz_MC_500_Nk_50.csv};
	%
\end{axis}
\end{tikzpicture}}%
		\label{fig:KL_Schwarz}}
	\hfil
	\subfloat[$K$ estimated with skew-Schwarz]
	{\resizebox{0.48\columnwidth}{!}{\begin{tikzpicture}
\begin{axis}[
	scaled ticks=false,
	tick label style={/pgf/number format/.cd},
	width = \figurewidth,
	height = \figureheight,
	xmin = 0,
	xmax = 0.08,
	ymax = 100,
	ymin = 0,
	ytick={0,10,20,30,40,50,60,70,80,90,100},
	yticklabels={$0$,$10$,$20$,$30$,$40$,$50$,$60$,$70$,$80$,$90$,$100$},
	xtick={0,0.01,0.02,0.03,0.04,0.05,0.06,0.07,0.08},
	xticklabels={$0$,$1$,$2$,$3$,$4$,$5$,$6$,$7$,$8$},
	grid,
	xlabel= outlieres $\epsilon$ in \%,
	ylabel= KL divergence,
	legend entries={{EM: skew-Gauß}\\
					{EM: skew-t, $\nu = 3$}\\
					{EM: skew-t, opt}\\
					{EM: skew-Huber}\\},
	legend pos=north west,
	legend cell align={left},
	cycle list name=matlabcolor
	]
	\addplot+[thick,matlabblue,line width=1.5pt] table[x index=0,y index=1,col sep=tab] {figures/t-opt/KL_BIC-Skew-Schwarz_MC_500_Nk_50.csv};
	\addplot+[thick,matlaborange,line width=1.5pt] table[x index=0,y index=5,col sep=tab] {figures/t-opt/KL_BIC-Skew-Schwarz_MC_500_Nk_50.csv};
	\addplot+[thick,matlabpurple,line width=1.5pt] table[x index=0,y index=3,col sep=tab] {figures/t-opt/KL_t_optBIC-Skew-Schwarz_MC_500_Nk_50.csv};
	\addplot+[thick,matlabyellow,line width=1.5pt] table[x index=0,y index=2,col sep=tab] {figures/t-opt/KL_BIC-Skew-Schwarz_MC_500_Nk_50.csv};
	%
\end{axis}
\end{tikzpicture}}%
		\label{fig:KL_Schwarz_skew}} \\
	\subfloat[$K = 3$ assumed to be known]
	{\resizebox{0.48\columnwidth}{!}{\begin{tikzpicture}
\begin{axis}[
	scaled ticks=false,
	tick label style={/pgf/number format/.cd},
	width = \figurewidth,
	height = \figureheight,
	xmin = 0,
	xmax = 0.08,
	ymax = 100,
	ymin = 0,
	ytick={0,10,20,30,40,50,60,70,80,90,100},
	yticklabels={$0$,$10$,$20$,$30$,$40$,$50$,$60$,$70$,$80$,$90$,$100$},
	xtick={0,0.01,0.02,0.03,0.04,0.05,0.06,0.07,0.08},
	xticklabels={$0$,$1$,$2$,$3$,$4$,$5$,$6$,$7$,$8$},
	grid,
	xlabel= outlieres $\epsilon$ in \%,
	ylabel= KL divergence,
	legend entries={{EM: Gauß}\\
					{EM: t, $\nu = 3$}\\
					{EM: t, opt}\\
					{EM: Huber}\\},
	legend pos=north west,
	legend cell align={left},
	cycle list name=matlabcolor
	]
	\addplot+[thick,matlabblue,line width=1.5pt] table[x index=0,y index=1,col sep=tab] {figures/t-opt/KL_True-K_MC_500_Nk_50.csv};
	\addplot+[thick,matlaborange,line width=1.5pt] table[x index=0,y index=5,col sep=tab] {figures/t-opt/KL_True-K_MC_500_Nk_50.csv};
	\addplot+[thick,matlabpurple,line width=1.5pt] table[x index=0,y index=3,col sep=tab] {figures/t-opt/KL_t_optTrue-K_MC_500_Nk_50.csv};
	\addplot+[thick,matlabyellow,line width=1.5pt] table[x index=0,y index=2,col sep=tab] {figures/t-opt/KL_True-K_MC_500_Nk_50.csv};
	%
\end{axis}
\end{tikzpicture}}%
		\label{fig:KL_true}}
	\hfil
	\subfloat[$K = 3$ assumed to be known]
	{\resizebox{0.48\columnwidth}{!}{\begin{tikzpicture}
\begin{axis}[
	scaled ticks=false,
	tick label style={/pgf/number format/.cd},
	width = \figurewidth,
	height = \figureheight,
	xmin = 0,
	xmax = 0.08,
	ymax = 100,
	ymin = 0,
	ytick={0,10,20,30,40,50,60,70,80,90,100},
	yticklabels={$0$,$10$,$20$,$30$,$40$,$50$,$60$,$70$,$80$,$90$,$100$},
	xtick={0,0.01,0.02,0.03,0.04,0.05,0.06,0.07,0.08},
	xticklabels={$0$,$1$,$2$,$3$,$4$,$5$,$6$,$7$,$8$},
	grid,
	xlabel= outlieres $\epsilon$ in \%,
	ylabel= KL divergence,
	legend entries={{EM: skew-Gauß}\\
					{EM: skew-t, $\nu = 3$}\\
					{EM: skew-t, opt}\\
					{EM: skew-Huber}\\},
	legend pos=north west,
	legend cell align={left},
	cycle list name=matlabcolor
	]
	\addplot+[thick,matlabblue,line width=1.5pt] table[x index=0,y index=1,col sep=tab] {figures/t-opt/KL_True-K-skew_MC_500_Nk_50.csv};
	\addplot+[thick,matlaborange,line width=1.5pt] table[x index=0,y index=5,col sep=tab] {figures/t-opt/KL_True-K-skew_MC_500_Nk_50.csv};
	\addplot+[thick,matlabpurple,line width=1.5pt] table[x index=0,y index=3,col sep=tab] {figures/t-opt/KL_t_optTrue-K-skew_MC_500_Nk_50.csv};
	\addplot+[thick,matlabyellow,line width=1.5pt] table[x index=0,y index=2,col sep=tab] {figures/t-opt/KL_True-K-skew_MC_500_Nk_50.csv};
	%
\end{axis}
\end{tikzpicture}}%
		\label{fig:KL_true_skew}}
	\caption{Kullback–Leibler divergences between estimated and true pdfs for $N_{K} = 50$, over different amounts of outliers. A lower KL divergence value indicates better performance. The data is generated as an epsilon contamination mixture consisting of a skew-Gaussian for the clustered data and a uniform distribution for the outliers.}
	\label{fig:KL}
\end{figure}
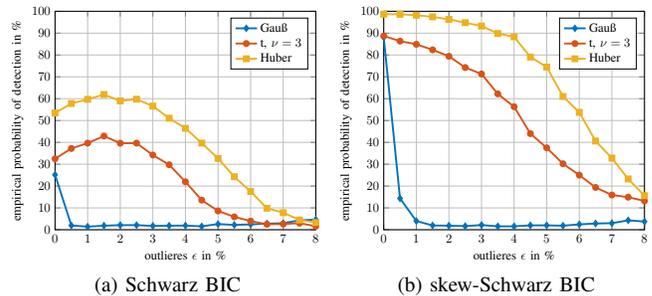
\begin{figure}[t]
	\centering
	\subfloat[Schwarz BIC] {\resizebox{0.48\columnwidth}{!}{\begin{tikzpicture}
	\begin{axis}[
	scaled ticks=false,
	tick label style={/pgf/number format/.cd},
	width = \figurewidth,
	height = \figureheight,
	xmin = 0,
	xmax = 0.08,
	ymax = 1,
	ymin = 0,
	ytick={0,0.1,0.2,0.3,0.4,0.5,0.6,0.7,0.8,0.9,1},
	yticklabels={$0$,$10$,$20$,$30$,$40$,$50$,$60$,$70$,$80$,$90$,$100$},
	xtick={0,0.01,0.02,0.03,0.04,0.05,0.06,0.07,0.08},
	xticklabels={$0$,$1$,$2$,$3$,$4$,$5$,$6$,$7$,$8$},
	grid,
	xlabel= outlieres $\epsilon$ in \%,
	ylabel= empirical probability of detection in \%,
	legend entries={{Gauß}\\
					{t, $\nu = 3$}\\
					{Huber}\\},
	legend pos=north east,
	legend cell align={left},
	cycle list name=matlabcolor
	]
	\plotfileNOlegend{figures/BIC_breakdown_BIC-Schwarz_MC_2000_Nk_50.csv}
	\end{axis}
\end{tikzpicture}}%
		\label{fig:schwarz-bic}}
	\hfil
	\subfloat[skew-Schwarz BIC] {\resizebox{0.48\columnwidth}{!}{\begin{tikzpicture}
	\begin{axis}[
	scaled ticks=false,
	tick label style={/pgf/number format/.cd},
	width = \figurewidth,
	height = \figureheight,
	xmin = 0,
	xmax = 0.08,
	ymax = 1,
	ymin = 0,
	ytick={0,0.1,0.2,0.3,0.4,0.5,0.6,0.7,0.8,0.9,1},
	yticklabels={$0$,$10$,$20$,$30$,$40$,$50$,$60$,$70$,$80$,$90$,$100$},
	xtick={0,0.01,0.02,0.03,0.04,0.05,0.06,0.07,0.08},
	xticklabels={$0$,$1$,$2$,$3$,$4$,$5$,$6$,$7$,$8$},
	grid,
	xlabel= outlieres $\epsilon$ in \%,
	ylabel= empirical probability of detection in \%,
	legend entries={{Gauß}\\
					{t, $\nu = 3$}\\
					{Huber}\\},
	legend pos=north east,
	legend cell align={left},
	cycle list name=matlabcolor
	]
	\plotfileNOlegend{figures/BIC_breakdown_BIC-Skew-Schwarz_MC_2000_Nk_50.csv}
	\end{axis}
\end{tikzpicture}}%
		\label{fig:skew-schwarz-bic}}
	\caption{Empirical probability of detection for the correct number of clusters over different amount of outliers, for  $N_{K} = 50$ samples per cluster. The data is generated as an epsilon contamination mixture consisting of a skew-Gaussian for the clustered data and a uniform distribution for the outliers. }
	\label{fig:data_31}
\end{figure}

\subsection{Results Based on Dataset 1: Robustness, Parameter Estimation and Clustering Accuracy}

Figure~\ref{fig:KL} shows the KL divergences over different amounts of outlier contamination, averaged over 500 Monte Carlo iterations. In the first row, the number of clusters is estimated with the Schwarz BIC from Eq.~\eqref{eqn:bic_schwarz} and in the second row the true number of clusters is assumed to be known. By comparing the first column (RES model) with the second column (RESK model), it becomes clear that, for this simulation, the proposed skew-Huber model performs better than all other RES and RESK models. This is the case, because of Huber's weight function $\psi(t_n)$ gives full weight to the data points with a squared distance smaller than $c^2$ while heavily down-weighting the outliers. One can also note that the skew-t with $\nu=3$ is outperformed in all cases, and even the oracle method skew-t, opt is outperformed for almost all cases when $K$ is assumed to be known. Until approximately 3\% of outliers, the KL divergences for the true and estimated number of clusters are almost constant and similar, because in this range the Schwarz BIC has a very high empirical probability to detect the correct number of clusters, as shown in Figure~\ref{fig:skew-schwarz-bic}. When using RES distributions, such a high empirical probability of detection cannot be reached, see Figure~\ref{fig:schwarz-bic}. As the amount of outliers increases and the performance of the Schwarz BIC starts to deteriorate, the KL divergence remains lower than for the case with the true number of clusters. This can be explained by the degree of freedom of opening up an additional cluster for the outliers, which leads to a deceptive better data fit, that is actually overfitting.

\begin{figure}[t]
	\centering
	\subfloat[Gauß]
	{\resizebox{0.48\columnwidth}{!}{\begin{tikzpicture}
	\begin{axis}[
	/pgf/number format/.cd,
	fixed,
	1000 sep={},
	width = \figurewidth,
	height = \figureheight,
	xmin = -15,
	xmax = 45,
	ymin = -20,
	ymax = 30,
	xlabel= Feature 1,
	ylabel= Feature 2,
	enlargelimits=false,
	colorbar, 
	point meta min=0,point meta max=400, 
	colorbar style={
		xshift = -2mm,
		ylabel = KL divergence,
		yticklabel style={
			text width=1em,
		}
	}
	]
	\addplot[contour prepared = {labels=false,},contour prepared format=matlab,line width=1.5pt] table[header=true] {figures/sensKL/sens_KL_EM_Gaus_ellip_Nk_50_step_5_MC_100_1.csv};
	\addplot[scatter/classes={1={matlabblue},2={matlaborange},3={matlabyellow}}, scatter, only marks, scatter src=explicit symbolic] table[x=x,y=y,meta=label] {figures/sensKL/sens_KL_EM_Gaus_ellip_Nk_50_step_5_MC_100_2.csv};   
	\end{axis}
\end{tikzpicture}}%
		\label{fig:sens_gaus}}
	\hfil
	\subfloat[skew-Gauß]
	{\resizebox{0.48\columnwidth}{!}{\begin{tikzpicture}
	\begin{axis}[
	/pgf/number format/.cd,
	fixed,
	1000 sep={},
	width = \figurewidth,
	height = \figureheight,
	xmin = -15,
	xmax = 45,
	ymin = -20,
	ymax = 30,
	xlabel= Feature 1,
	ylabel= Feature 2,
	colorbar, 
	point meta min=0,point meta max=400, 
	colorbar style={
		xshift = -2mm,
		ylabel = KL divergence,
		yticklabel style={
			text width=1em,
		}
	}
	]
	\addplot[contour prepared = {labels=false,},contour prepared format=matlab,line width=1.5pt] table[header=true] {figures/sensKL/sens_KL_EM_Gaus_skew_Nk_50_step_5_MC_100_1.csv};
	\addplot[scatter/classes={1={matlabblue},2={matlaborange},3={matlabyellow}}, scatter, only marks, scatter src=explicit symbolic] table[x=x,y=y,meta=label] {figures/sensKL/sens_KL_EM_Gaus_skew_Nk_50_step_5_MC_100_2.csv};   
	\end{axis}
\end{tikzpicture}}%
		\label{fig:sens_skew_gaus}} \\
	\subfloat[t]
	{\resizebox{0.48\columnwidth}{!}{\begin{tikzpicture}
	\begin{axis}[
	/pgf/number format/.cd,
	fixed,
	1000 sep={},
	width = \figurewidth,
	height = \figureheight,
	xmin = -15,
	xmax = 45,
	ymin = -20,
	ymax = 30,
	xlabel= Feature 1,
	ylabel= Feature 2,
	colorbar, 
	point meta min=0,point meta max=400, 
	colorbar style={
		xshift = -2mm,
		ylabel = KL divergence,
		yticklabel style={
			text width=1em,
		}
	}
	]
	\addplot[contour prepared = {labels=false,},contour prepared format=matlab,line width=1.5pt] table[header=true] {figures/sensKL/sens_KL_EM_t_ellip_Nk_50_step_5_MC_100_1.csv};
	\addplot[scatter/classes={1={matlabblue},2={matlaborange},3={matlabyellow}}, scatter, only marks, scatter src=explicit symbolic] table[x=x,y=y,meta=label] {figures/sensKL/sens_KL_EM_t_ellip_Nk_50_step_5_MC_100_2.csv};   
	\end{axis}
\end{tikzpicture}}%
		\label{fig:sens_t}}
	\hfil
	\subfloat[skew-t]
	{\resizebox{0.48\columnwidth}{!}{\begin{tikzpicture}
	\begin{axis}[
	/pgf/number format/.cd,
	fixed,
	1000 sep={},
	width = \figurewidth,
	height = \figureheight,
	xmin = -15,
	xmax = 45,
	ymin = -20,
	ymax = 30,
	xlabel= Feature 1,
	ylabel= Feature 2,
	colorbar, 
	point meta min=0,point meta max=400, 
	colorbar style={
		xshift = -2mm,
		ylabel = KL divergence,
		yticklabel style={
			text width=1em,
		}
	}
	]
	\addplot[contour prepared = {labels=false,},contour prepared format=matlab,line width=1.5pt] table[header=true] {figures/sensKL/sens_KL_EM_t_skew_Nk_50_step_5_MC_100_1.csv};
	\addplot[scatter/classes={1={matlabblue},2={matlaborange},3={matlabyellow}}, scatter, only marks, scatter src=explicit symbolic] table[x=x,y=y,meta=label] {figures/sensKL/sens_KL_EM_t_skew_Nk_50_step_5_MC_100_2.csv};   
	\end{axis}
\end{tikzpicture}}%
		\label{fig:sens_skew_t}} \\
	\subfloat[Huber]
	{\resizebox{0.48\columnwidth}{!}{\begin{tikzpicture}
	\begin{axis}[
	/pgf/number format/.cd,
	fixed,
	1000 sep={},
	width = \figurewidth,
	height = \figureheight,
	xmin = -15,
	xmax = 45,
	ymin = -20,
	ymax = 30,
	xlabel= Feature 1,
	ylabel= Feature 2,
	colorbar, 
	point meta min=0,point meta max=400, 
	colorbar style={
		xshift = -2mm,
		ylabel = KL divergence,
		yticklabel style={
			text width=1em,
		}
	}
	]
	\addplot[contour prepared = {labels=false,},contour prepared format=matlab,line width=1.5pt] table[header=true] {figures/sensKL/sens_KL_EM_Huber_ellip_Nk_50_step_5_MC_100_1.csv};
	\addplot[scatter/classes={1={matlabblue},2={matlaborange},3={matlabyellow}}, scatter, only marks, scatter src=explicit symbolic] table[x=x,y=y,meta=label] {figures/sensKL/sens_KL_EM_Huber_ellip_Nk_50_step_5_MC_100_2.csv};   
	\end{axis}
\end{tikzpicture}}%
		\label{fig:sens_huber}}
	\hfil
	\subfloat[skew-Huber]
	{\resizebox{0.48\columnwidth}{!}{\begin{tikzpicture}
	\begin{axis}[
	/pgf/number format/.cd,
	fixed,
	1000 sep={},
	width = \figurewidth,
	height = \figureheight,
	xmin = -15,
	xmax = 45,
	ymin = -20,
	ymax = 30,
	xlabel= Feature 1,
	ylabel= Feature 2,
	colorbar, 
	point meta min=0,point meta max=400, 
	colorbar style={
		xshift = -2mm,
		ylabel = KL divergence,
		yticklabel style={
			text width=1em,
		}
	}
	]
	\addplot[contour prepared = {labels=false,},contour prepared format=matlab,line width=1.5pt] table[header=true] {figures/sensKL/sens_KL_EM_Huber_skew_Nk_50_step_5_MC_100_1.csv};
	\addplot[scatter/classes={1={matlabblue},2={matlaborange},3={matlabyellow}}, scatter, only marks, scatter src=explicit symbolic] table[x=x,y=y,meta=label] {figures/sensKL/sens_KL_EM_Huber_skew_Nk_50_step_5_MC_100_2.csv};   
	\end{axis}
\end{tikzpicture}}%
		\label{fig:sens_skew_huber}}
	\caption{Sensitivity curves for $N_{K} = 50$ that show six exemplary results for the KL divergence, while estimating $K$ with the Schwarz/skew-Schwarz BIC, as a function of the single replacement outlier position. A lower KL divergence value indicates better performance. The clusters are generated from skew-Gaussian distributions and one randomly chosen sample is replaced with an outlier from the range [-15, 45], [-20, 30] in x, y-dimension.}
	\label{fig:sens_all}
\end{figure}
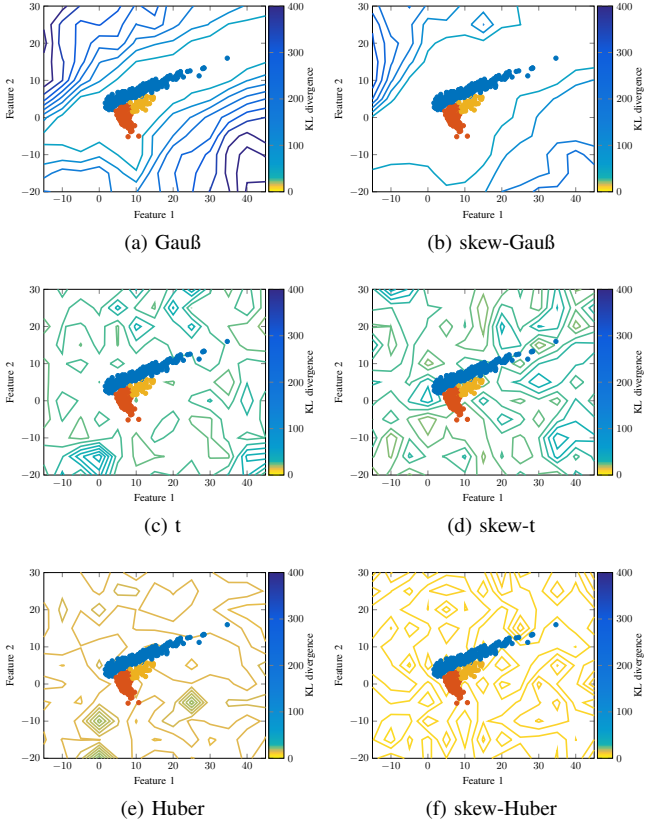

In Figure~\ref{fig:sens_all}, six exemplary sensitivity curves are shown. The number of clusters is estimated with the Schwarz/skew-Schwarz BIC. As desired for a robust estimator, for this example, the skew-Huber estimator has the lowest KL divergence. It is almost not influenced by the position of the outlier, as indicated by the almost constant KL divergence over the grid. For the Gauß and skew-Gauß estimators, the performance quickly deteriorates as soon as the outlier is positioned outside the clusters. The other robust estimators, i.e. the t, skew-t and Huber show a stable but inferior performance compared to the skew-Huber estimator.

Tables~\ref{tb:conf_0} and \ref{tb:conf_2} show the confusion matrices with 0\% and 2\% of outliers. In the 0\% case, all methods have a very high probability of assigning the data points to the correct cluster, with the best result for both Huber (RES and RESK) estimators. For 2\% of outlier contamination, both Gaussian estimators break down, whereas the t and Huber based estimators are able to maintain the high performance from the 0\% case, again, with the best result for the skew-Huber distribution based estimator.
\begin{table*}[ht]
	\centering
	\hfill
	\caption{Confusion matrices for simulated data with $\epsilon = 0\%$}
	\label{tb:conf_0}
	\begin{tabu}{cc ccc|ccc|ccc|ccc|ccc|ccc}
		\toprule \multicolumn{2}{c}{$\epsilon = 0\%$} & \multicolumn{3}{c}{Gauß} & \multicolumn{3}{c}{t} & \multicolumn{3}{c}{Huber} & \multicolumn{3}{c}{skew-Gauß} & \multicolumn{3}{c}{skew-t}& \multicolumn{3}{c}{skew-Huber}\\ \cmidrule{3-20} 
		& & 1 & 2 & 3 & 1 & 2 & 3 & 1 & 2 & 3 & 1 & 2 & 3 & 1 & 2 & 3 & 1 & 2 & 3 \\ \cmidrule{2-20} 
		\multirow{3}{*}{\begin{sideways}true\end{sideways}} & 1 & \textbf{87.2} &  1.5 &  11.3 & \textbf{95.6} &  2.6 & 1.8 &\textbf{95.4} &  1.9 &   2.7 & \textbf{94.9} &  1 & 4.1 & \textbf{93} & 1.5 &5.5 & \textbf{97.4} &  1.4 &  1.2\\
		& 2 &  0.1  & \textbf{98.8} &  1.1 & 0  & \textbf{99.8} &  0.2 & 0  & \textbf{99.7} &  0.3 & 0.1  & \textbf{98.9} &  1 & 0 & \textbf{99.8} & 0.2 &0  & \textbf{99.9} &  0.1\\
		& 3 &  3.2 & 1.1 & \textbf{95.7} & 3.7  & 2 & \textbf{94.3}  & 3  & 1.9 & \textbf{95.1}  & 1.9  & 1.3 & \textbf{96.8} & 3.1 & 4.6 & \textbf{92.3} & 2.1  & 3.5 & \textbf{94.4} \\ \cmidrule{2-20}
		& \O & \multicolumn{3}{c}{93.9} & \multicolumn{3}{c}{96.5} & \multicolumn{3}{c}{96.7} & \multicolumn{3}{c}{96.9} & \multicolumn{3}{c}{95.1}& \multicolumn{3}{c}{97.2}\\\bottomrule
	\end{tabu}
	\hfill
\end{table*}
\begin{table*}[ht]
	\centering
	\hfill
	\caption{Confusion matrices for simulated data with $\epsilon = 2\%$}
	\label{tb:conf_2}
	\begin{tabu}{cc ccc|ccc|ccc|ccc|ccc|ccc}
		\toprule \multicolumn{2}{c}{$\epsilon = 2\%$} & \multicolumn{3}{c}{Gauß} & \multicolumn{3}{c}{t} & \multicolumn{3}{c}{Huber} & \multicolumn{3}{c}{skew-Gauß} & \multicolumn{3}{c}{skew-t}& \multicolumn{3}{c}{skew-Huber}\\ \cmidrule{3-20}
		& & 1 & 2 & 3 & 1 & 2 & 3 & 1 & 2 & 3 & 1 & 2 & 3 & 1 & 2 & 3 & 1 & 2 & 3 \\ \cmidrule{2-20} 
		\multirow{3}{*}{\begin{sideways}true\end{sideways}} & 1 & \textbf{96} &  2.3 &  1.7 & \textbf{94.9} &  2.8 &   2.3 &\textbf{94.2} &  1.9 &   3.9 & \textbf{98.1} &  1.2 & 0.7 & \textbf{89} & 1.7 & 9.3 & \textbf{96.4} &  1.4 &  2.2\\
		& 2 &  1.1  & \textbf{98.5} &  0.4 & 0  & \textbf{99.8} &  0.2 & 0.1  & \textbf{99.6} &  0.3 & 1.9  & \textbf{97.7} &  0.4 & 0.1 & \textbf{99.7} & 0.2 &0  & \textbf{99.8} &  0.2\\
		& 3 &  83.3 & 3.8 & \textbf{12.9} &  4.1  & 2 & \textbf{93.9}  &3.7  & 1.9 & \textbf{94.4}  & 76.9  & 11.5 & \textbf{11.6} & 4.5 & 5 & \textbf{90.5} & 2.4  & 3.4 & \textbf{94.2} \\ \cmidrule{2-20}
		& \O & \multicolumn{3}{c}{69.1} & \multicolumn{3}{c}{96.1} & \multicolumn{3}{c}{96} & \multicolumn{3}{c}{69.1} & \multicolumn{3}{c}{93.1}& \multicolumn{3}{c}{96.8}\\\bottomrule
	\end{tabu}
	\hfill
\end{table*}

\subsection{Results Based on Dataset 2: Assessment of Bias and Convergence Speed}
In this section, we examine the bias behavior and convergence speed based on Dataset 2. Because this data follows a skew-t distribution with $\nu=3$, the skew-t, $\nu=3$ is the MLE for $\epsilon=0 \%$ outliers. In Figure \ref{fig:KL_conv}, we show the KL divergence over different amounts of outliers. As expected, the best performance can be observed for the skew-t estimator, closely followed by the skew-Huber estimator. As it can be seen from Figure \ref{fig:KL_conv_true_skew}, the KL divergence nearly approaches zero for $\epsilon=0 \%$, which indicates, in combination with the low variance, that the bias of the algorithms is very small.

To asses the convergence speed of the EM algorithm, the average number of EM iterations is plotted in Figure~\ref{fig:RL_conv}. Two main observations can be made. Firstly, the RESK-based EM algorithm takes much longer to converge than the RES algorithm and, additionally, it has a much more varying stopping time. The anomaly of the skew-Gaussian based EM stopping faster for lager amount of outliers, can be probably explained by the fact, that the algorithm completely breaks down and stops prematurely at a local optimum.

Lastly, to further investigate the bias behavior and convergence speed, the results of the EM algorithms are shown for different amounts of samples per cluster. No outliers are included in this case. As shown in Figure \ref{fig:KL_nk}, the larger the amount of samples per cluster, the better the estimation of the true pdf and the lower the variance in the estimation. For a very large number of samples per cluster, the estimated and true pdf are almost identical. This confirms, for the given example, that the approximate ML estimate is near-optimal if the nominal model that is used in the EM algorithm and the data generating model match exactly. For the the RESK algorithms, the convergence speed of the EM algorithms increases for larger sample sizes, as shown in Figure \ref{fig:RL_nk}.
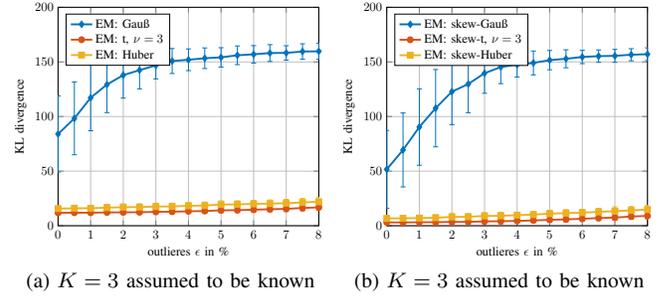
\begin{figure}[t]
	\centering
	\subfloat[$K = 3$ assumed to be known]
	{\resizebox{0.48\columnwidth}{!}{\begin{tikzpicture}
\begin{axis}[
	scaled ticks=false,
	tick label style={/pgf/number format/.cd},
	width = \figurewidth,
	height = \figureheight,
	xmin = 0,
	xmax = 0.08,
	ymax = 200,
	ymin = 0,
	xtick={0,0.01,0.02,0.03,0.04,0.05,0.06,0.07,0.08},
	xticklabels={$0$,$1$,$2$,$3$,$4$,$5$,$6$,$7$,$8$},
	grid,
	xlabel= outlieres $\epsilon$ in \%,
	ylabel= KL divergence,
	legend entries={{EM: Gauß}\\
		{EM: t, $\nu = 3$}\\
		{EM: Huber}\\},
	legend pos=north west,
	legend cell align={left},
	cycle list name=matlabcolor
	]
	\addplot+[thick,matlabblue,line width=1.5pt,error bars/.cd, y dir=both, y explicit] table[x index=0,y index=1,col sep=tab, y error index=4] {figures/conv/KL_True-K_MC_500_Nk_200.csv};
	\addplot+[thick,matlaborange,line width=1.5pt,error bars/.cd, y dir=both, y explicit] table[x index=0,y index=2,col sep=tab, y error index=5] {figures/conv/KL_True-K_MC_500_Nk_200.csv};
	\addplot+[thick,matlabyellow,line width=1.5pt,error bars/.cd, y dir=both, y explicit] table[x index=0,y index=3,col sep=tab, y error index=6] {figures/conv/KL_True-K_MC_500_Nk_200.csv};
\end{axis}
\end{tikzpicture}}%
		\label{fig:KL_conv_true}}
	\hfil
	\subfloat[$K = 3$ assumed to be known]
	{\resizebox{0.48\columnwidth}{!}{\begin{tikzpicture}
\begin{axis}[
	scaled ticks=false,
	tick label style={/pgf/number format/.cd},
	width = \figurewidth,
	height = \figureheight,
	xmin = 0,
	xmax = 0.08,
	ymax = 200,
	ymin = 0,
	xtick={0,0.01,0.02,0.03,0.04,0.05,0.06,0.07,0.08},
	xticklabels={$0$,$1$,$2$,$3$,$4$,$5$,$6$,$7$,$8$},
	grid,
	xlabel= outlieres $\epsilon$ in \%,
	ylabel= KL divergence,
	legend entries={{EM: skew-Gauß}\\
		{EM: skew-t, $\nu = 3$}\\
		{EM: skew-Huber}\\},
	legend pos=north west,
	legend cell align={left},
	cycle list name=matlabcolor
	]
	\addplot+[thick,matlabblue,line width=1.5pt,error bars/.cd, y dir=both, y explicit] table[x index=0,y index=1,col sep=tab, y error index=4] {figures/conv/KL_True-K-skew_MC_500_Nk_200.csv};
	\addplot+[thick,matlaborange,line width=1.5pt,error bars/.cd, y dir=both, y explicit] table[x index=0,y index=2,col sep=tab, y error index=5] {figures/conv/KL_True-K-skew_MC_500_Nk_200.csv};
	\addplot+[thick,matlabyellow,line width=1.5pt,error bars/.cd, y dir=both, y explicit] table[x index=0,y index=3,col sep=tab, y error index=6] {figures/conv/KL_True-K-skew_MC_500_Nk_200.csv};
\end{axis}
\end{tikzpicture}}%
		\label{fig:KL_conv_true_skew}}
	\caption{Kullback-Leibler divergences between estimated and true pdfs for $N_{K} = 200$, over different amounts of outliers. A lower KL divergence value indicates better performance. The data is generated as an epsilon contamination mixture consisting of a skew-t with $\nu=3$ for the clustered data and a uniform distribution for the outliers. }
	\label{fig:KL_conv}
\end{figure}
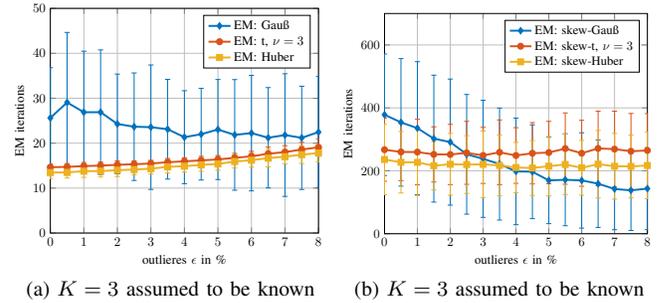
\begin{figure}[t]
	\centering
	\subfloat[$K = 3$ assumed to be known]
	{\resizebox{0.48\columnwidth}{!}{\begin{tikzpicture}
\begin{axis}[
	scaled ticks=false,
	tick label style={/pgf/number format/.cd},
	width = \figurewidth,
	height = \figureheight,
	xmin = 0,
	xmax = 0.08,
	ymax = 50,
	ymin = 0,
	xtick={0,0.01,0.02,0.03,0.04,0.05,0.06,0.07,0.08},
	xticklabels={$0$,$1$,$2$,$3$,$4$,$5$,$6$,$7$,$8$},
	grid,
	xlabel= outlieres $\epsilon$ in \%,
	ylabel= EM iterations,
	legend entries={{EM: Gauß}\\
		{EM: t, $\nu = 3$}\\
		{EM: Huber}\\},
	legend pos=north east,
	legend cell align={left},
	cycle list name=matlabcolor
	]
	\addplot+[thick,matlabblue,line width=1.5pt,error bars/.cd, y dir=both, y explicit] table[x index=0,y index=1,col sep=tab, y error index=4] {figures/conv/RL_True-K_MC_500_Nk_200.csv};
	\addplot+[thick,matlaborange,line width=1.5pt,error bars/.cd, y dir=both, y explicit] table[x index=0,y index=2,col sep=tab, y error index=5] {figures/conv/RL_True-K_MC_500_Nk_200.csv};
	\addplot+[thick,matlabyellow,line width=1.5pt,error bars/.cd, y dir=both, y explicit] table[x index=0,y index=3,col sep=tab, y error index=6] {figures/conv/RL_True-K_MC_500_Nk_200.csv};
	%
\end{axis}
\end{tikzpicture}}%
		\label{fig:RL_conv_true}}
	\hfil
	\subfloat[$K = 3$ assumed to be known]
	{\resizebox{0.48\columnwidth}{!}{\begin{tikzpicture}
\begin{axis}[
	scaled ticks=false,
	tick label style={/pgf/number format/.cd},
	width = \figurewidth,
	height = \figureheight,
	xmin = 0,
	xmax = 0.08,
	ymax = 700,
	ymin = 0,
	xtick={0,0.01,0.02,0.03,0.04,0.05,0.06,0.07,0.08},
	xticklabels={$0$,$1$,$2$,$3$,$4$,$5$,$6$,$7$,$8$},
	grid,
	xlabel= outlieres $\epsilon$ in \%,
	ylabel= EM iterations,
	legend entries={{EM: skew-Gauß}\\
		{EM: skew-t, $\nu = 3$}\\
		{EM: skew-Huber}\\},
	legend pos=north east,
	legend cell align={left},
	cycle list name=matlabcolor
	]
	\addplot+[thick,matlabblue,line width=1.5pt,error bars/.cd, y dir=both, y explicit] table[x index=0,y index=1,col sep=tab, y error index=4] {figures/conv/RL_True-K-skew_MC_500_Nk_200.csv};
	\addplot+[thick,matlaborange,line width=1.5pt,error bars/.cd, y dir=both, y explicit] table[x index=0,y index=2,col sep=tab, y error index=5] {figures/conv/RL_True-K-skew_MC_500_Nk_200.csv};
	\addplot+[thick,matlabyellow,line width=1.5pt,error bars/.cd, y dir=both, y explicit] table[x index=0,y index=3,col sep=tab, y error index=6] {figures/conv/RL_True-K-skew_MC_500_Nk_200.csv};
	%
\end{axis}
\end{tikzpicture}}%
		\label{fig:RL_conv_true_skew}}
	\caption{EM iterations for $N_{K} = 200$, over different amounts of outliers. Less iterations indicate a better performance. The data is generated as an epsilon contamination mixture consisting of a skew-t with $\nu=3$ for the clustered data and a uniform distribution for the outliers. }
	\label{fig:RL_conv}
\end{figure}
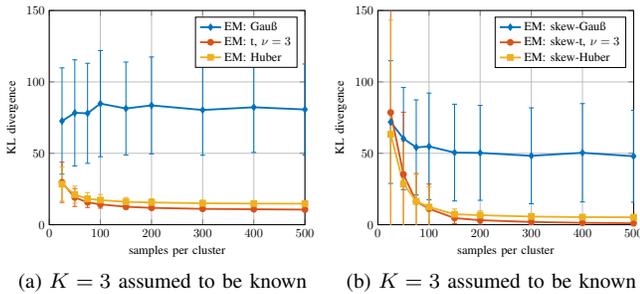
\begin{figure}[t]
	\centering
	\subfloat[$K = 3$ assumed to be known]
	{\resizebox{0.48\columnwidth}{!}{\begin{tikzpicture}
\begin{axis}[
	scaled ticks=false,
	tick label style={/pgf/number format/.cd},
	width = \figurewidth,
	height = \figureheight,
	xmin = 0,
	xmax = 500,
	ymax = 150,
	ymin = 0,
	grid,
	xlabel= samples per cluster,
	ylabel= KL divergence,
	legend entries={{EM: Gauß}\\
		{EM: t, $\nu = 3$}\\
		{EM: Huber}\\},
	legend pos=north east,
	legend cell align={left},
	cycle list name=matlabcolor
	]
	\addplot+[thick,matlabblue,line width=1.5pt,error bars/.cd, y dir=both, y explicit] table[x index=0,y index=1,col sep=tab, y error index=4] {figures/nk/KL_True-K_MC_500_eps_0.csv};
	\addplot+[thick,matlaborange,line width=1.5pt,error bars/.cd, y dir=both, y explicit] table[x index=0,y index=2,col sep=tab, y error index=5] {figures/nk/KL_True-K_MC_500_eps_0.csv};
	\addplot+[thick,matlabyellow,line width=1.5pt,error bars/.cd, y dir=both, y explicit] table[x index=0,y index=3,col sep=tab, y error index=6] {figures/nk/KL_True-K_MC_500_eps_0.csv};
	%
\end{axis}
\end{tikzpicture}}%
		\label{fig:KL_nk_true}}
	\hfil
	\subfloat[$K = 3$ assumed to be known]
	{\resizebox{0.48\columnwidth}{!}{\begin{tikzpicture}
\begin{axis}[
	scaled ticks=false,
	tick label style={/pgf/number format/.cd},
	width = \figurewidth,
	height = \figureheight,
	xmin = 0,
	xmax = 500,
	ymax = 150,
	ymin = 0,
	grid,
	xlabel= samples per cluster,
	ylabel= KL divergence,
	legend entries={{EM: skew-Gauß}\\
		{EM: skew-t, $\nu = 3$}\\
		{EM: skew-Huber}\\},
	legend pos=north east,
	legend cell align={left},
	cycle list name=matlabcolor
	]
	\addplot+[thick,matlabblue,line width=1.5pt,error bars/.cd, y dir=both, y explicit] table[x index=0,y index=1,col sep=tab, y error index=4] {figures/nk/KL_True-K-skew_MC_500_eps_0.csv};
	\addplot+[thick,matlaborange,line width=1.5pt,error bars/.cd, y dir=both, y explicit] table[x index=0,y index=2,col sep=tab, y error index=5] {figures/nk/KL_True-K-skew_MC_500_eps_0.csv};
	\addplot+[thick,matlabyellow,line width=1.5pt,error bars/.cd, y dir=both, y explicit] table[x index=0,y index=3,col sep=tab, y error index=6] {figures/nk/KL_True-K-skew_MC_500_eps_0.csv};
	%
\end{axis}
\end{tikzpicture}}%
		\label{fig:KL_nk_true_skew}}
	\caption{Kullback–Leibler divergences between estimated and true pdfs, over different samples per cluster. A lower KL divergence value indicates better performance. The data is generated as a skew-t with $\nu=3$ for the clustered data without any outlier contamination. }
	\label{fig:KL_nk}
\end{figure}
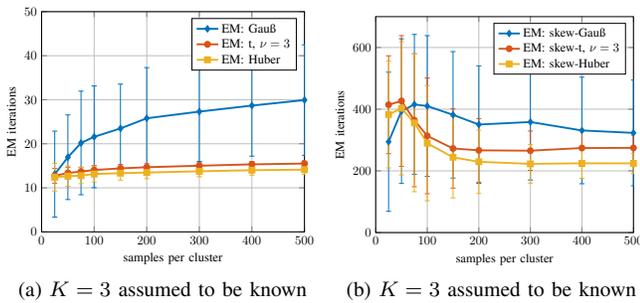
\begin{figure}[t]
	\centering
	\subfloat[$K = 3$ assumed to be known]
	{\resizebox{0.48\columnwidth}{!}{\begin{tikzpicture}
\begin{axis}[
	scaled ticks=false,
	tick label style={/pgf/number format/.cd},
	width = \figurewidth,
	height = \figureheight,
	xmin = 0,
	xmax = 500,
	ymax = 50,
	ymin = 0,
	grid,
	xlabel= samples per cluster,
	ylabel= EM iterations,
	legend entries={{EM: Gauß}\\
		{EM: t, $\nu = 3$}\\
		{EM: Huber}\\},
	legend pos=north east,
	legend cell align={left},
	cycle list name=matlabcolor
	]
	\addplot+[thick,matlabblue,line width=1.5pt,error bars/.cd, y dir=both, y explicit] table[x index=0,y index=1,col sep=tab, y error index=4] {figures/nk/RL_True-K_MC_500_eps_0.csv};
	\addplot+[thick,matlaborange,line width=1.5pt,error bars/.cd, y dir=both, y explicit] table[x index=0,y index=2,col sep=tab, y error index=5] {figures/nk/RL_True-K_MC_500_eps_0.csv};
	\addplot+[thick,matlabyellow,line width=1.5pt,error bars/.cd, y dir=both, y explicit] table[x index=0,y index=3,col sep=tab, y error index=6] {figures/nk/RL_True-K_MC_500_eps_0.csv};
	%
\end{axis}
\end{tikzpicture}}%
		\label{fig:RL_nk_true}}
	\hfil
	\subfloat[$K = 3$ assumed to be known]
	{\resizebox{0.48\columnwidth}{!}{\begin{tikzpicture}
\begin{axis}[
	scaled ticks=false,
	tick label style={/pgf/number format/.cd},
	width = \figurewidth,
	height = \figureheight,
	xmin = 0,
	xmax = 500,
	ymax = 700,
	ymin = 0,
	grid,
	xlabel= samples per cluster,
	ylabel= EM iterations,
	legend entries={{EM: skew-Gauß}\\
		{EM: skew-t, $\nu = 3$}\\
		{EM: skew-Huber}\\},
	legend pos=north east,
	legend cell align={left},
	cycle list name=matlabcolor
	]
	\addplot+[thick,matlabblue,line width=1.5pt,error bars/.cd, y dir=both, y explicit] table[x index=0,y index=1,col sep=tab, y error index=4] {figures/nk/RL_True-K-skew_MC_500_eps_0.csv};
	\addplot+[thick,matlaborange,line width=1.5pt,error bars/.cd, y dir=both, y explicit] table[x index=0,y index=2,col sep=tab, y error index=5] {figures/nk/RL_True-K-skew_MC_500_eps_0.csv};
	\addplot+[thick,matlabyellow,line width=1.5pt,error bars/.cd, y dir=both, y explicit] table[x index=0,y index=3,col sep=tab, y error index=6] {figures/nk/RL_True-K-skew_MC_500_eps_0.csv};
	%
\end{axis}
\end{tikzpicture}}%
		\label{fig:RL_nk_true_skew}}
	\caption{EM iterations for $\epsilon = 0\%$, over different samples per cluster. Less iterations indicate a better performance. The data is generated as a skew-t with $\nu=3$ for the clustered data without any outlier contamination. }
	\label{fig:RL_nk}
\end{figure}

\subsection{Real Data Simulations}
\subsubsection{Wine Data}
The wine quality data set from the UCI Machine Learning Repository (\url{https://archive.ics.uci.edu/ml/datasets/wine+quality}), is composed of 12 different attributes. To distinguish between red and white wine samples, we chose four attributes: volatile acidity, residual sugar, chlorides and total sulfur dioxide, from the data set. The first three attributes are visualized in Figure~\ref{fig:data_red_white_wine}. The number of clusters, i.e. $K = 2$ is assumed to be known and, to increase the difficulty, $\epsilon = 3\%$ of uniformly distributed replacement outliers are added in every dimension. From Table~\ref{tb:conf_real_data}, it is observed that the modeling of skewness increases the probability of assigning a data point to the correct cluster. Again, the skew-Huber shows the best results.
\begin{table*}[t]
	\centering
	\hfill
	\caption{Confusion matrices for the wine data set.}
	\label{tb:conf_real_data}
	\begin{tabu}{cc cc|cc|cc|cc|cc|cc}
		\toprule \multicolumn{2}{c}{$\epsilon = 3\%$} & \multicolumn{2}{c}{Gauß} & \multicolumn{2}{c}{t} & \multicolumn{2}{c}{Huber} & \multicolumn{2}{c}{skew-Gauß} & \multicolumn{2}{c}{skew-t}& \multicolumn{2}{c}{skew-Huber}\\ \cmidrule{3-14}
		& & 1 & 2 & 1 & 2 & 1 & 2 & 1 & 2 & 1 & 2 & 1 & 2 \\ \cmidrule{2-14} 
		\multirow{2}{*}{\begin{sideways}true\end{sideways}} & 1 & \textbf{97.8} &  2.2 & \textbf{87.1} &  12.9  &\textbf{67} &  33 & \textbf{97.8} &  2.2 & \textbf{93.6} & 6.4 & \textbf{93.7} & 6.3 \\
		& 2 &  94.8  & \textbf{5.2}  & 18  & \textbf{82} & 5.6  & \textbf{94.4}  & 94.8  & \textbf{5.2}& 11.5 & \textbf{88.5}  &2.8  & \textbf{97.2} \\ \cmidrule{2-14} 
		& \O & \multicolumn{2}{c}{51.5} & \multicolumn{2}{c}{84.5} & \multicolumn{2}{c}{80.7} & \multicolumn{2}{c}{51.5} & \multicolumn{2}{c}{90.9}& \multicolumn{2}{c}{95.5}\\\bottomrule
	\end{tabu}
	\hfill
\end{table*}

\subsubsection{Crabs Data}
The crab data set was first used by \cite{Campbell.1974} and can be downloaded at \url{https://doi.org/10.24097/wolfram.70344.data}. It is composed of five different length measurements on Leptograpsus crabs, of which all five were used to distinguish between blue/orange and male/female crabs. The number of clusters $K = 4$ is assumed to be known and no additional outliers are added. Table~\ref{tb:conf_crab} and \ref{tb:conf_crab_skew} show, that the modeling of skewness, again, increases the overall performance. For this data set, the skew-t distribution based estimator showed best overall performance.
\begin{table*}[t]
	\centering
	\hfill
	\caption{Confusion matrices for the crab data set.}
	\label{tb:conf_crab}
	\begin{tabu}{ccc cccc|cccc|cccc}
		\toprule \multicolumn{3}{c}{} & \multicolumn{4}{c}{Gauß} & \multicolumn{4}{c}{t} & \multicolumn{4}{c}{Huber} \\ \cmidrule{4-15}
		& & & 1 & 2 & 3 & 4 & 1 & 2 & 3 & 4 & 1 & 2 & 3 & 4 \\ \cmidrule{2-15} 
		\multirow{2}{*}{blue} & male & 1 & \textbf{33.7} &  14.9 & 25.5  & 25.9 & \textbf{90} & 10 & 0 & 0 & \textbf{38.2} & 18.5 & 26.2 & 17.1 \\
		& female & 2 &  30  & \textbf{23.7} & 34.3 & 12 & 76.9 & \textbf{23.1} & 0 & 0 & 7 & \textbf{72.4} & 10.4 & 10.2\\
		\multirow{2}{*}{orange} & male & 3 & 26 &  11.2  & \textbf{38.8} & 24 & 0 & 10 & \textbf{40} & 50 & 19.9 & 10.4 & \textbf{40.8} & 28.9\\
		& female & 4 & 40.4 & 5.2 & 16.4 &  \textbf{38}  & 0 & 5.5 & 34.5 & \textbf{60} & 37.6 & 7.8 & 17.1 & \textbf{37.5} \\ \cmidrule{2-15}
		& \O & & \multicolumn{4}{c}{33.5} & \multicolumn{4}{c}{53.3} & \multicolumn{4}{c}{47.2}\\\bottomrule
	\end{tabu}
	\hfill
\end{table*}
\begin{table*}[t]
	\centering
	\hfill
	\caption{Confusion matrices for the crab data set.}
	\label{tb:conf_crab_skew}
	\begin{tabu}{ccc cccc|cccc|cccc}
		\toprule \multicolumn{3}{c}{} & \multicolumn{4}{c}{skew-Gauß} & \multicolumn{4}{c}{skew-t} & \multicolumn{4}{c}{skew-Huber} \\ \cmidrule{4-15}
		& & & 1 & 2 & 3 & 4 & 1 & 2 & 3 & 4 & 1 & 2 & 3 & 4 \\ \cmidrule{2-15} 
		\multirow{2}{*}{blue} & male & 1 & \textbf{9.2} &  24.4 & 59.9 & 6.5 & \textbf{0.4} & 27 & 72.6 & 0 & \textbf{36.9} & 24 & 39.1 & 0 \\
		& female & 2 &  0  & \textbf{88} & 0.1 & 11.9 & 0 & \textbf{100} & 0 & 0 & 0.1 & \textbf{94.3} & 0 & 5.6\\
		\multirow{2}{*}{orange} & male & 3 & 6.4 &  11.1 & \textbf{82.5} & 0 & 11.4 & 0 & \textbf{88.6} & 0 & 39.2 & 9.4 & \textbf{51.4} & 0\\
		& female & 4 & 9.3 & 20.3 & 0.1 &  \textbf{70.3}  & 7.9 & 0.8 & 6.3 & \textbf{85} & 4.1 & 18.7 & 4.6 & \textbf{72.6} \\ \cmidrule{2-15}
		& \O & & \multicolumn{4}{c}{62.5} & \multicolumn{4}{c}{68.5} & \multicolumn{4}{c}{63.8}\\\bottomrule
	\end{tabu}
	\hfill
\end{table*}

\section{Conclusion}
\label{sec:conclusion}
We have presented a finite mixture model based on the proposed family of RESK distributions which is capable of integrating robustness and skewness into a single framework. Special attention was given to a newly proposed skew-Huber distribution, which has the attractive property that resulting estimators give maximal weight to data points with a distance smaller than a chosen threshold value while heavily down weighting outliers. The down weighting is asymmetric by taking skewness of the distribution into account. For the proposed family of RESK distributions, we derived an EM algorithm to estimate the cluster parameters and memberships. The performance was evaluated numerically in terms of parameter estimation and clustering performance for skewed and heavy-tailed mixture models, robustness against outliers, and convergence speed in terms of the number of required EM iterations. The proposed methods demonstrated promising performance on all simulated and real-world examples. This showcases their usefulness in simultaneously addressing skewness, outliers and heavy tailed data. The provided framework, therefore, could be useful in a variety of cluster analysis applications where the data may be skewed and heavy-tailed. Future work may consider deriving new robust cluster enumeration criteria and further robust M-estimators to cluster skewed data with outliers. New asymmetric M-estimators can be designed by varying the density generator within the class of RESK distributions according to the user's preference in terms of weighting the data points. Future work may also include deriving a stochastic representation for the skew-Huber distribution or other M-estimators. Based on this stochastic representation, the corresponding approximate ML estimator could be replaced by the exact ML estimator.

\appendices
\section{Derivation of M-Step Estimates}
\label{sec:app}
Using the matrix calculus rules from \cite{Magnus.2007, Abadir.2005}, $\bF$ is a differentiable $m \times p$ matrix function of a $n \times q$ matrix $\bX$. Then, the Jacobian matrix of $\bF$ at $\bX$ is a $mp \times nq$ matrix
\begin{equation}
	D\bF(\bX) = \frac{\partial \vecop(\bF(\bX))}{\partial (\vecop(\bX))^{\top}}.
\end{equation}
Starting with the differential of the complete data log-likelihood regarding $\bxi_{m}$, we define $\bF$ as a $1 \times 1$ scalar function of the $r \times 1$ vector $\bxi_{m}$. Hence, the resulting Jacobian matrix is just the gradient and is of size $1 \times r$. Setting $\bF$ equal to \eqref{eqn:complete}, i.e., 
\begin{align}
	\bF(\bxi_{m}) =& \sum_{n=1}^{N} \sum_{m=1}^{l} v_{nm} \left(\ln(\gamma_{m}) + \ln(f_{\text{s}}(\bx_{n} | \btheta_{m}))\right)
\end{align}
and applying the differential leads to
\begin{align}
	\text{d}\bF&(\bxi_{m}) = \sum_{n=1}^{N} v_{nm} \text{ d} \left(\ln(f_{\text{s}}(\bx_{n} | \btheta_{m}))\right)\notag \\
	=& \sum_{n = 1}^{N} \frac{ 2 \left| \bOmega_{m}^{-1} \right|^{\frac{1}{2}}}{f_{\text{s}}(\bx_{n} | \btheta_{m})} \left[\text{d}g\left( \underline{t}_{nm}\right)F\left(\kappa_{nm}\right) + g\left( \underline{t}_{nm}\right) \text{d}F\left(\kappa_{nm}\right)\right].
	\label{eqn:dFxi}
\end{align}
Both differentials are calculated separately as
\begin{align}
	\text{d} g\left( \underline{t}_{nm}\right) =& - g^{\prime}\left( \underline{t}_{nm}\right) 2 \left(\bx_{n} - \bxi_{m}\right)^{\top} \bOmega_{m}^{-1} \text{d}\bxi_{m}\notag \\
	\begin{split}
		=&  g\left( \underline{t}_{nm}\right) \psi\left(\underline{t}_{nm}\right) 2 
		\\&\times \left(\left(\bx_{n} - \bxi_{m}\right)^{\top} \bS_{m}^{-1} - \eta_{nm} \blambda_{m}^{\top} \bS_{m}^{-1}\right) \text{d}\bxi_{m}
	\end{split}
\end{align}
with
\begin{equation}
	g'\left( t_{nm}\right) = - \psi\left(t_{nm}\right) g\left( t_{nm}\right),
	\label{eqn:gprime}
\end{equation}
the Sherman-Morrison formula \cite{Bartlett.1951}
\begin{equation}
	\left(\bA + \bb \bc^{\top}\right)^{-1} = \bA^{-1} - \frac{\bA^{-1}\bb \bc^{\top}\bA^{-1}}{1 + \bc^{\top}\bA^{-1}\bb}
	\label{eqnA:Sherman}
\end{equation}
and
\begin{align}
	\text{d}F\left(\kappa_{nm}\right)& \notag \\
	=& F^{\prime}\left(\kappa_{nm}\right) \text{d}\left(\frac{\eta_{nm}}{\tau_{m}}\sqrt{2 \psi(\underline{t}_{nm})}\right)\notag \\
	\begin{split}
		=& - F^{\prime}\left(\kappa_{nm}\right) \Biggl[\frac{ \blambda_{m}^{\top} \bS_{m}^{-1}\text{d}\bxi_{m}}{\sqrt{1 + \blambda_{m}^{\top}\bS_{m}^{-1}\blambda_{m}}} \sqrt{2 \psi(\underline{t}_{nm})} 
		\\&+ 2 \frac{\eta_{nm}}{\tau_{m}} \frac{\eta(\underline{t}_{nm}) }{\sqrt{2 \psi(\underline{t}_{nm})}} \left(\bx_{n} - \bxi_{m}\right)^{\top} \bOmega_{m}^{-1} \text{d}\bxi_{m} \Biggr]
	\end{split} \notag \\
	\begin{split}
		=& F\left(\kappa_{nm}\right) \Psi\left(\kappa_{nm}\right) \Biggl[ \tau_{m} \sqrt{2 \psi(\underline{t}_{nm})} \blambda_{m}^{\top} \bS_{m}^{-1} 
		\\&+ 2 \frac{\eta_{nm}}{\tau_{m}} \frac{\eta(\underline{t}_{nm}) }{\sqrt{2 \psi(\underline{t}_{nm})}} \left(\bx_{n} - \bxi_{m}\right)^{\top} \bS_{m}^{-1} 
		\\&- 2 \frac{\eta_{nm}^{2}}{\tau_{m}} \frac{\eta(\underline{t}_{nm}) }{\sqrt{2 \psi(\underline{t}_{nm})}}  \blambda_{m}^{\top} \bS_{m}^{-1}\Biggr]\text{d}\bxi_{m}
	\end{split}
\end{align}
with
\begin{equation}
	F^{\prime}(x)  = - \Psi(x) F(x).
\end{equation}
Using the scalar values
\begin{align}
	e_{0,nm} = 2 \psi\left(\underline{t}_{nm}\right) + 2 \Psi\left(\kappa_{nm}\right)\frac{\eta_{nm}}{\tau_{m}} \frac{\eta(\underline{t}_{nm}) }{\sqrt{2 \psi(\underline{t}_{nm})}},
	\label{eqn:e0}
\end{align}
\begin{align}
	\begin{split}
		e_{1,nm} =& 2 \psi\left(\underline{t}_{nm}\right)\eta_{nm} - \Psi\left(\kappa_{nm}\right) \tau_{m} \sqrt{2 \psi(\underline{t}_{nm})} 
		\\&+  \Psi\left(\kappa_{nm}\right) 2 \frac{\eta_{nm}^{2}}{\tau_{m}} \frac{\eta(\underline{t}_{nm}) }{\sqrt{2 \psi(\underline{t}_{nm})}}
	\end{split}
	\label{eqn:e1}
\end{align}
the differentials can be combined to the gradient as follows
\begin{align}
	\text{D}\bF(\bxi_{m}) =& \sum_{n = 1}^{N} v_{nm} \left(e_{0,nm} \left(\bx_{n} - \bxi_{m}\right)^{\top}  - e_{1,nm} \blambda_{m}^{\top}\right) \bS_{m}^{-1}.
	\label{eqn:grad_xi}
\end{align}
Setting Equation~\eqref{eqn:grad_xi} to zero and solving for $\bxi_{m}$ leads to the ML estimate. As discussed in Section~\ref{ap:EM_skew}, finding an exact solution, is not feasible for general RESK distributions. Therefore, the values $e_{0,nm}$ and $e_{1,nm}$, are replaced by their approximations, i.e. $\tilde{e}_{0,nm}$, $\tilde{e}_{1,nm}$. These approximations are constants with respect to the unknown parameters $\btheta_{m}$. They are introduced by approximating $\underline{t}_{nm}$, $\tau_{m}$, $\eta_{nm}$, and $\kappa_{nm}$ with their corresponding estimates from the previous iteration of Algorithm~\ref{alg:em_skew}. The resulting approximate ML estimate is the obtained by setting the approximate gradient to zero:
\begin{align}
	& \sum_{n = 1}^{N} v_{nm} \left(\tilde{e}_{0,nm} \left(\bx_{n} - \bxi_{m}\right)^{\top}- \tilde{e}_{1,nm} \blambda_{m}^{\top} \right)\bS_{m}^{-1} \overset{!}{=} 0  \notag\\
	\Rightarrow & \sum_{n = 1}^{N} v_{nm} \left( \tilde{e}_{0,nm} \left(\bx_{n} - \bxi_{m}\right) - \tilde{e}_{1,nm}  \blambda_{m}\right) = 0 \notag\\
	\Rightarrow & \bxihat_{m} = \frac{\sum_{n = 1}^{N} v_{nm} \left(\tilde{e}_{0,nm} \bx_{n}  - \tilde{e}_{1,nm}  \blambda_{m}\right)}{\sum_{n = 1}^{N} v_{nm} \tilde{e}_{0,nm} }.
\end{align}
Now, $\bF$ is defined as a $1 \times 1$ scalar function of the $r \times 1$ vector $\blambda_{m}$. Hence, the resulting Jacobian matrix is just the gradient and is of size $1 \times r$. Setting $\bF$ equal to \eqref{eqn:emml_skew} and applying the differential gives
\begin{align}
	\text{d}\bF(\blambda_{m}) =& \sum_{n=1}^{N} v_{nm} \text{ d} \left(\ln(f_{\text{s}}(\bx_{n} | \btheta_{m}))\right)\notag \\
	\begin{split}
		=& \sum_{n = 1}^{N} \frac{2}{f_{\text{s}}(\bx_{n} | \btheta_{m})} 
		\Bigl[ \text{d}\left(\left| \bOmega_{m}\right|^{-\frac{1}{2}}\right) g\left( \underline{t}_{nm}\right) F\left(\kappa_{nm}\right) 
		\\&+ \left| \bOmega_{m} \right|^{-\frac{1}{2}}\text{d} g\left( \underline{t}_{nm}\right) F\left(\kappa_{nm}\right) 
		\\&+ \left| \bOmega_{m} \right|^{-\frac{1}{2}} g\left( \underline{t}_{nm}\right) \text{d}F\left(\kappa_{nm}\right) \Bigr].
	\end{split}
\end{align}
Again, the occurring differentials are calculated separately. Firstly,
\begin{align}
	\text{d}(|\bOmega_{m}&|^{-\frac{1}{2}}) \notag\\
	=& -\frac{1}{2} \left| \bOmega_{m} \right|^{-\frac{1}{2}} \left| \bOmega_{m} \right|^{-1} \left| \bOmega_{m} \right| \Tr\left(\bOmega_{m}^{-1} \text{d}\bOmega_{m}\right)\notag \\
	\begin{split}
		=& -\frac{1}{2} \left| \bOmega_{m} \right|^{-\frac{1}{2}} 
		\\& \times \Tr\left(\left(\bS_{m}^{-1} - \frac{\bS_{m}^{-1} \blambda_{m} \blambda_{m}^{\top} \bS_{m}^{-1}}{1 + \blambda_{m}^{\top}\bS_{m}^{-1}\blambda_{m}}\right) \text{d} \left(\blambda_{m}\blambda_{m}^{\top}\right) \right)
	\end{split} \notag \\
	\begin{split}
		=& -\frac{1}{2} \left| \bOmega_{m} \right|^{-\frac{1}{2}} \Bigl[\vecop\left(\bS_{m}^{-1}\right)^{\top} 
		\\&- \tau_{m}^{2} \vecop\left( \bS_{m}^{-1} \blambda_{m} \blambda_{m}^{\top} \bS_{m}^{-1}\right)^{\top} \Bigr] \text{d} \vecop\left(\blambda_{m}\blambda_{m}^{\top}\right)
	\end{split}\notag \\
	\begin{split}
		=& -\frac{1}{2} \left| \bOmega_{m} \right|^{-\frac{1}{2}} \Bigl[\vecop\left(\bS_{m}^{-1}\blambda_{m}\right)^{\top} + \vecop\left(\blambda_{m}^{\top} \bS_{m}^{-1}\right)^{\top}  
		\\&- \tau_{m}^{2} \Bigl[ \blambda_{m}^{\top} \bS_{m}^{-1}\blambda_{m}  \otimes \blambda_{m}^{\top} \bS_{m}^{-1}
		\\&+  \blambda_{m}^{\top} \bS_{m}^{-1}  \otimes \blambda_{m}^{\top} \bS_{m}^{-1}\blambda_{m}\Bigr]\Bigr]\text{d} \blambda_{m}
	\end{split}\notag \\
	=& -\left| \bOmega_{m} \right|^{-\frac{1}{2}} \tau_{m}^{2} \blambda_{m}^{\top} \bS_{m}^{-1}\text{d}\blambda_{m},
	\label{eqn:d_det_omega}
\end{align}
with
\begin{equation}
	\text{d}\vecop\left(\blambda_{m}\blambda_{m}^{\top}\right) = \left(\left(\blambda_{m} \otimes \bI_{r}\right)+ \left(\bI_{r}\otimes \blambda_{m} \right)\right) \text{d}\blambda_{m}.
\end{equation}
Secondly, with $\bxt_{n} = \bx_{n} - \bxi_{m}$
\begin{align}
	\text{d} g\left( \underline{t}_{nm}\right) =& g^{\prime} \left( \underline{t}_{nm}\right) \bxt_{n}^{\top} \text{d}\left(\bOmega_{m}^{-1}\right) \bxt_{n}\notag \\
	\begin{split}
		=& - g^{\prime}\left( \underline{t}_{nm}\right) \bxt_{n}^{\top} \left(\bS_{m} + \blambda_{m}\blambda_{m}^{\top}\right)^{-1} 
		\\& \times \text{ d}\left(\bS_{m} + \blambda_{m}\blambda_{m}^{\top}\right) \left(\bS_{m} + \blambda_{m}\blambda_{m}^{\top}\right)^{-1}\bxt_{n}
	\end{split}\notag \\
	\begin{split}
		=& - g^{\prime}\left( \underline{t}_{nm}\right) \left(\bxt_{n}^{\top}\bS_{m}^{-1} - \eta_{nm} \blambda_{m}^{\top}\bS_{m}^{-1} \right) 
		\\& \times \text{ d}\left(\blambda_{m}\blambda_{m}^{\top}\right)  \left(\bxt_{n}^{\top}\bS_{m}^{-1} - \eta_{nm} \blambda_{m}^{\top}\bS_{m}^{-1} \right)
	\end{split}
\end{align}
with vectorization
\begin{align}
	\text{d} \vecop &\left(g\left( \underline{t}_{nm}\right)\right) \notag \\
	\begin{split}
		=&- g^{\prime}\left( \underline{t}_{nm}\right) \Bigl[\left(\bxt_{n}^{\top}\bS_{m}^{-1} - \eta_{nm} \blambda_{m}^{\top}\bS_{m}^{-1} \right)
		\\& \otimes  \left(\bxt_{n}^{\top}\bS_{m}^{-1} - \eta_{nm} \blambda_{m}^{\top}\bS_{m}^{-1} \right)\Bigr] \text{d}\vecop\left(\blambda_{m}\blambda_{m}^{\top}\right) 
	\end{split} \notag \\
	=& g\left( \underline{t}_{nm}\right) \psi\left(t_{nm}\right) 2 \left(\eta_{nm} \bxt_{n}^{\top}\bS_{m}^{-1} - \eta_{nm}^{2} \blambda_{m}^{\top}\bS_{m}^{-1}\right)\text{d}\blambda_{m}
	\label{eqn:d_t_lambda_vec}
\end{align}
and thirdly
\begin{align}
	\text{d}F&\left(\kappa_{nm}\right) \notag \\
	=& F^{\prime}\left(\kappa_{nm}\right) \text{d}\left(\frac{\blambda_{m}^{\top} \bS_{m}^{-1}\bxt_{n}}{\sqrt{1 + \blambda_{m}^{\top}\bS_{m}^{-1}\blambda_{m}}}\sqrt{2 \psi(\underline{t}_{nm})}\right)\notag \\
	\begin{split}
		=& F^{\prime}\left(\kappa_{nm}\right) \Bigl[\tau_{m} \sqrt{2 \psi(\underline{t}_{nm})} \text{d}\left(\blambda_{m}^{\top} \right)\bS_{m}^{-1}\bxt_{n} 
		\\&- \frac{\eta_{nm}}{\tau_{m}^{2}} \sqrt{2 \psi(\underline{t}_{nm})} \tau_{m}^{2} \frac{\text{d}\left(1 + \blambda_{m}^{\top}\bS_{m}^{-1}\blambda_{m}\right)}{2 \sqrt{1 + \blambda_{m}^{\top}\bS_{m}^{-1}\blambda_{m}}} 
		\\&+ \frac{\eta_{nm}}{\tau_{m}} \frac{\eta(\underline{t}_{nm})}{\sqrt{2 \psi(\underline{t}_{nm})}} \bxt_{n}^{\top} \text{d}\left(\bOmega_{m}^{-1}\right) \bxt_{n}\Bigr]
	\end{split}\notag \\
	\begin{split}
		=& F^{\prime}\left(\kappa_{nm}\right) \Bigl[\tau_{m} \sqrt{2 \psi(\underline{t}_{nm})} \text{d}\left(\blambda_{m}^{\top} \right)\bS_{m}^{-1}\bxt_{n} 
		\\&- \frac{1}{2} \eta_{nm}\tau_{m}\sqrt{2 \psi(\underline{t}_{nm})} \text{d}\left( \blambda_{m}^{\top} \bS_{m}^{-1}\blambda_{m}\right) 
		\\&- \frac{\eta_{nm}}{\tau_{m}} \frac{\eta(\underline{t}_{nm})}{\sqrt{2 \psi(\underline{t}_{nm})}} \left(\bxt_{n}^{\top}\bS_{m}^{-1} - \eta_{nm} \blambda_{m}^{\top}\bS_{m}^{-1} \right) 
		\\& \times \text{d}\left(\blambda_{m}\blambda_{m}^{\top}\right)  \left(\bS_{m}^{-1}\bxt_{n} - \eta_{nm} \bS_{m}^{-1}\blambda_{m}\right)\Bigr]
	\end{split}
\end{align}
with vectorization
\begin{align}
	\text{d}\vecop&\left(F\left(\kappa_{nm}\right)\right) \notag \\
	\begin{split}
		=& F^{\prime}\left(\kappa_{nm}\right) \Biggl[\tau_{m}\sqrt{2 \psi(\underline{t}_{nm})} \left(\bxt_{n}^{\top}\bS_{m}^{-1} \otimes \bI_{1}\right) \text{d}\blambda_{m}^{\top}
		\\&- \frac{1}{2} \eta_{nm}\tau_{m}\sqrt{2 \psi(\underline{t}_{nm})} \text{d}\vecop\left( \blambda_{m}^{\top} \bS_{m}^{-1}\blambda_{m}\right)
		\\& - \frac{\eta_{nm}}{\tau_{m}} \frac{\eta(\underline{t}_{nm})}{\sqrt{2 \psi(\underline{t}_{nm})}} \Bigl(\left(\bxt_{n}^{\top}\bS_{m}^{-1} - \eta_{nm} \blambda_{m}^{\top}\bS_{m}^{-1} \right) 
		\\&\otimes \left(\bxt_{n}^{\top}\bS_{m}^{-1} - \eta_{nm} \blambda_{m}^{\top}\bS_{m}^{-1} \right)\Bigr)\text{d}\vecop\left(\blambda_{m}\blambda_{m}^{\top}\right) \Biggr]
	\end{split}\notag \\
	\begin{split}
		=& - F\left(\kappa_{nm}\right) \Psi\left(\kappa_{nm}\right)\Biggl[\tau_{m}\sqrt{2 \psi(\underline{t}_{nm})} \bxt_{n}^{\top}\bS_{m}^{-1}  
		\\&- \eta_{nm}\tau_{m} \sqrt{2 \psi(\underline{t}_{nm})}\blambda_{m}^{\top}\bS_{m}^{-1}
		- \frac{\eta_{nm}^{2}}{\tau_{m}} \frac{2 \eta(\underline{t}_{nm})}{\sqrt{2 \psi(\underline{t}_{nm})}}  \bxt_{n}^{\top}\bS_{m}^{-1} 
		\\&+ \frac{\eta_{nm}^{3}}{\tau_{m}} \frac{2 \eta(\underline{t}_{nm})}{\sqrt{2 \psi(\underline{t}_{nm})}} \blambda_{n}^{\top}\bS_{m}^{-1} \Biggr]\text{d}\blambda_{m}.
	\end{split}
\end{align}
Combining the differentials with
\begin{align}
	\begin{split}
		e_{2,nm} =&  \tau_{m}^{2} + \psi\left(\underline{t}_{nm}\right) 2 \eta_{nm}^{2} - \Psi\left(\kappa_{nm}\right)\eta_{nm}\tau_{m} \sqrt{2 \psi(\underline{t}_{nm})} 
		\\&+ \Psi\left(\kappa_{nm}\right) 2 \frac{\eta_{nm}^{3}}{\tau_{m}} \frac{\eta(\underline{t}_{nm})}{\sqrt{2 \psi(\underline{t}_{nm})}}
	\end{split}
	\label{eqn:e2}
\end{align}
the gradient becomes
\begin{align}
	\text{D}\bF(\blambda_{m}) =& \sum_{n = 1}^{N} v_{nm} \left( e_{1,nm}  \bxt_{n}^{\top}\bS_{m}^{-1} - e_{2,nm}\blambda_{m}^{\top} \bS_{m}^{-1} \right).
	\label{eqn:grad_lambda}
\end{align}
Analogously to Eq.~\eqref{eqn:grad_xi}, the gradient in Eq.~\eqref{eqn:grad_lambda} is approximated by replacing $e_{1,nm}$ and $e_{2,nm}$ by their approximations, i.e. $\tilde{e}_{1,nm}$, $\tilde{e}_{2,nm}$. Then, the approximate ML estimate follows by setting the approximate gradient to zero and solving for~$\blambda_{m}$:
\begin{align}
	\blambdahat_{m} = \frac{\sum_{n = 1}^{N} v_{nm} \tilde{e}_{1,nm} \bxt_{n}}{\sum_{n = 1}^{N} v_{nm} \tilde{e}_{2,nm} } .
\end{align}

Lastly, $\bF$ is defined as a $1 \times 1$ scalar function of the $r \times r$ matrix $\bS_{m}$. Hence, the resulting Jacobian matrix is just the gradient and is of size $1 \times r^2$. Setting $\bF$ equal to \eqref{eqn:emml_skew} and applying the differential
\begin{align}
	\text{d}\bF(\bS_{m}) =& \sum_{n=1}^{N} v_{nm} \text{ d} \left(\ln(f_{\text{s}}(\bx_{n} | \btheta_{m}))\right)\notag \\
	\begin{split}
		=& \sum_{n = 1}^{N} \frac{2}{f_{\text{s}}(\bx_{n} | \btheta_{m})} 
		\Bigl[ \text{d}\left(\left| \bOmega_{m}^{-1} \right|^{\frac{1}{2}}\right) g\left( \underline{t}_{nm}\right) F\left(\kappa_{nm}\right) 
		\\&+ \left| \bOmega_{m}^{-1} \right|^{\frac{1}{2}}\text{d} g\left( \underline{t}_{nm}\right) F\left(\kappa_{nm}\right)
		\\&+ \left| \bOmega_{m}^{-1} \right|^{\frac{1}{2}} g\left( \underline{t}_{nm}\right) \text{d}F\left(\kappa_{nm}\right) \Bigr],
	\end{split}
\end{align}
again, leads to three differentials that are calculated separately. Starting with
\begin{align}
	\text{d}(| \bOmega_{m} &|^{-\frac{1}{2}}) \notag \\
	=& -\frac{1}{2} \left| \bOmega_{m} \right|^{-\frac{1}{2}} \left| \bOmega_{m} \right|^{-1} \left| \bOmega_{m} \right| \Tr\left(\bOmega_{m}^{-1} \text{d}\bOmega_{m}\right)\notag \\
	\begin{split}
		=& -\frac{1}{2} \left| \bOmega_{m} \right|^{-\frac{1}{2}} \Tr\Bigl[\bS_{m}^{-1}\text{d} \left(\bS_{m}\right) 
		\\&- \tau_{m}^{2} \bS_{m}^{-1} \blambda_{m} \blambda_{m}^{\top} \bS_{m}^{-1} \text{d} \left(\bS_{m}\right) \Bigr]
	\end{split} \notag \\
	\begin{split}
		=&  -\frac{1}{2} \left| \bOmega_{m}\right|^{-\frac{1}{2}} \Bigl[\vecop\left(\bS_{m}^{-1}\right)^{\top}
		\\&- \tau_{m}^{2} \vecop\left( \bS_{m}^{-1} \blambda_{m} \blambda_{m}^{\top} \bS_{m}^{-1}\right)^{\top}\Bigr]  \text{d} \vecop\left(\bS_{m}\right)
	\end{split} \notag \\
	\begin{split}
		=& -\frac{1}{2} \left| \bOmega_{m}\right|^{-\frac{1}{2}} \Bigl[\vecop\left(\bS_{m}^{-1}\right)^{\top} 
		\\&- \tau_{m}^{2} \left( \blambda_{m}^{\top} \bS_{m}^{-1}  \otimes \blambda_{m}^{\top}\bS_{m}^{-1}  \right) \Bigr]\bD_{r} \text{ d}\vechop\left(\bS_{m}\right)
	\end{split}
	\label{eqn:dvec_lnomega_sigma}
\end{align}
followed by
\begin{align}
	\text{d} g\left( \underline{t}_{nm}\right) =& g^{\prime} \left( \underline{t}_{nm}\right) \bxt_{n}^{\top} \text{d}\left(\bOmega_{m}^{-1}\right) \bxt_{n}\notag \\
	\begin{split}
		=& - g^{\prime}\left( \underline{t}_{nm}\right) \bxt_{n}^{\top} \left(\bS_{m} + \blambda_{m}\blambda_{m}^{\top}\right)^{-1} 
		\\& \times \text{d}\left(\bS_{m}\right) \left(\bS_{m} + \blambda_{m}\blambda_{m}^{\top}\right)^{-1}\bxt_{n}
	\end{split}\notag \\
	\begin{split}
		=& - g^{\prime}\left( \underline{t}_{nm}\right) \left(\bxt_{n}^{\top}\bS_{m}^{-1} - \eta_{nm} \blambda_{m}^{\top}\bS_{m}^{-1} \right) 
		\\& \times \text{ d}\left(\bS_{m}\right)  \left(\bxt_{n}^{\top}\bS_{m}^{-1} - \eta_{nm} \blambda_{m}^{\top}\bS_{m}^{-1} \right)
	\end{split}
\end{align}
with the vectorization
\begin{align}
	\text{d} \vecop&\left(g\left( \underline{t}_{nm}\right)\right) \notag \\
	\begin{split}
		=& - g^{\prime}\left( \underline{t}_{nm}\right) \Bigl[\left(\bxt_{n}^{\top}\bS_{m}^{-1} - \eta_{nm} \blambda_{m}^{\top}\bS_{m}^{-1} \right) 
		\\& \otimes  \left(\bxt_{n}^{\top}\bS_{m}^{-1} - \eta_{nm} \blambda_{m}^{\top}\bS_{m}^{-1} \right)\Bigr]  \text{ d}\vecop\left(\bS_{m}\right) 
	\end{split} \notag \\
	\begin{split}
		=& g\left( \underline{t}_{nm}\right) \psi\left( \underline{t}_{nm}\right) \Bigl[\left(\bxt_{n}^{\top}\bS_{m}^{-1} \otimes \bxt_{n}^{\top}\bS_{m}^{-1}\right) 
		\\&- \left(\bxt_{n}^{\top}\bS_{m}^{-1} \otimes \eta_{nm} \blambda_{m}^{\top}\bS_{m}^{-1}\right) 
		- \left(\eta_{nm} \blambda_{m}^{\top}\bS_{m}^{-1} \otimes \bxt_{n}^{\top}\bS_{m}^{-1}\right) 
		\\&+ \left(\eta_{nm}^{2} \blambda_{m}^{\top}\bS_{m}^{-1} \otimes  \blambda_{m}^{\top}\bS_{m}^{-1}\right)\Bigr] \bD_{r} \text{ d}\vechop\left(\bS_{m}\right)
	\end{split}
	\label{eqn:dvec_t_sigma}
\end{align}
and finally
\begin{align}
	\text{d}F\left(\kappa_{nm}\right) =& F^{\prime}\left(\kappa_{nm}\right) \text{d}\left(\frac{\blambda_{m}^{\top} \bS_{m}^{-1}\left(\bx_{n} - \bxi_{m}\right)}{\sqrt{1 + \blambda_{m}^{\top}\bS_{m}^{-1}\blambda_{m}}} \sqrt{2 \psi(\underline{t}_{nm})} \right)\notag \\
	\begin{split}
		=& F^{\prime}\left(\kappa_{nm}\right) \Biggl[\tau_{m}  \sqrt{2 \psi(\underline{t}_{nm})} \blambda_{m}^{\top} \text{d}\left(\bS_{m}^{-1} \right)\bxt_{n} 
		\\&- \frac{\eta_{nm}}{\tau_{m}^{2}} \sqrt{2 \psi(\underline{t}_{nm})} \tau_{m}^{2} \frac{ \blambda_{m}^{\top} \text{d}\left(\bS_{m}^{-1}\right)\blambda_{m}}{2 \sqrt{1 + \blambda_{m}^{\top}\bS_{m}^{-1}\blambda_{m}}} 
		\\&+ \frac{\eta_{nm}}{\tau_{m}} \frac{\eta(\underline{t}_{nm})}{\sqrt{2 \psi(\underline{t}_{nm})}} \bxt_{n}^{\top} \text{d}\left(\bOmega_{m}^{-1} \right)\bxt_{n} \Biggr]
	\end{split} \notag \\
	\begin{split}
		=& F^{\prime}\left(\kappa_{nm}\right) \Biggl[-\tau_{m} \sqrt{2 \psi(\underline{t}_{nm})} \blambda_{m}^{\top} \bS_{m}^{-1} \text{d}\left(\bS_{m}\right) \bS_{m}^{-1}\bxt_{n} 
		\\&+ \frac{1}{2} \eta_{nm}\tau_{m} \sqrt{2 \psi(\underline{t}_{nm})}  \blambda_{m}^{\top} \bS_{m}^{-1} \text{d}\left(\bS_{m}\right) \bS_{m}^{-1}\blambda_{m}
		\\& - \frac{\eta_{nm}}{\tau_{m}} \frac{\eta(\underline{t}_{nm})}{\sqrt{2 \psi(\underline{t}_{nm})}} \left(\bxt_{n}^{\top}\bS_{m}^{-1} - \eta_{nm} \blambda_{m}^{\top}\bS_{m}^{-1} \right) 
		\\& \times \text{d}\left(\bS_{m}\right)  \left(\bS_{m}^{-1} \bxt_{n}- \eta_{nm} \bS_{m}^{-1} \blambda_{m}\right)\Biggr]
	\end{split}
\end{align}
with vectorization
\begin{align}
	\begin{split}
		\text{d}\vecop&\left(F\left(\kappa_{nm}\right)\right) 
		\\=& F^{\prime}\left(\kappa_{nm}\right) \Biggl[ - \tau_{m} \sqrt{2 \psi(\underline{t}_{nm})} \left(\bxt_{n}^{\top}\bS_{m}^{-1} \otimes \blambda_{m}^{\top}\bS_{m}^{-1}\right)
		\\& + \frac{1}{2} \eta_{nm}\tau_{m} \sqrt{2 \psi(\underline{t}_{nm})} \left(\blambda_{m}^{\top}\bS_{m}^{-1} \otimes \blambda_{m}^{\top}\bS_{m}^{-1}\right) 
		\\&- \frac{\eta_{nm}}{\tau_{m}} \frac{\eta(\underline{t}_{nm})}{\sqrt{2 \psi(\underline{t}_{nm})}} \Bigl[\left(\bxt_{n}^{\top}\bS_{m}^{-1} \otimes \bxt_{n}^{\top}\bS_{m}^{-1}\right) 
		\\&- \left(\bxt_{n}^{\top}\bS_{m}^{-1} \otimes \eta_{nm} \blambda_{m}^{\top}\bS_{m}^{-1}\right) - \left(\eta_{nm} \blambda_{m}^{\top}\bS_{m}^{-1} \otimes \bxt_{n}^{\top}\bS_{m}^{-1}\right) 
		\\&+ \left(\eta_{nm}^{2} \blambda_{m}^{\top}\bS_{m}^{-1} \otimes  \blambda_{m}^{\top}\bS_{m}^{-1}\right)\Bigr]\Biggr]\bD_{r}\text{d}\vechop\left( \bS_{m}\right)
	\end{split} \notag\\
	\begin{split}
		=& F\left(\kappa_{nm}\right) \Psi\left(\kappa_{nm}\right) \Biggl[\frac{\tau_{m}}{2}  \sqrt{2 \psi(\underline{t}_{nm})} \left(\blambda_{m}^{\top}\bS_{m}^{-1} \otimes \bxt_{n}^{\top}\bS_{m}^{-1}\right) 
		\\&+ \frac{\tau_{m}}{2}  \sqrt{2 \psi(\underline{t}_{nm})} \left(\bxt_{n}^{\top}\bS_{m}^{-1} \otimes \blambda_{m}^{\top}\bS_{m}^{-1}\right) 
		\\&- \frac{1}{2} \eta_{nm}\tau_{m}  \sqrt{2 \psi(\underline{t}_{nm})} \left(\blambda_{m}^{\top}\bS_{m}^{-1} \otimes \blambda_{m}^{\top}\bS_{m}^{-1}\right)
		\\&+ \frac{\eta_{nm}}{\tau_{m}} \frac{\eta(\underline{t}_{nm})}{\sqrt{2 \psi(\underline{t}_{nm})}} \left(\bxt_{n}^{\top}\bS_{m}^{-1} \otimes \bxt_{n}^{\top}\bS_{m}^{-1}\right) 
		\\&- \frac{\eta_{nm}^{2}}{\tau_{m}} \frac{\eta(\underline{t}_{nm})}{\sqrt{2 \psi(\underline{t}_{nm})}} \left(\bxt_{n}^{\top}\bS_{m}^{-1} \otimes  \blambda_{m}^{\top}\bS_{m}^{-1}\right) 
		\\&- \frac{\eta_{nm}^{2}}{\tau_{m}} \frac{\eta(\underline{t}_{nm})}{\sqrt{2 \psi(\underline{t}_{nm})}} \left( \blambda_{m}^{\top}\bS_{m}^{-1} \otimes \bxt_{n}^{\top}\bS_{m}^{-1}\right) 
		\\&+ \frac{\eta_{nm}^{3}}{\tau_{m}} \frac{\eta(\underline{t}_{nm})}{\sqrt{2 \psi(\underline{t}_{nm})}} \left( \blambda_{m}^{\top}\bS_{m}^{-1} \otimes  \blambda_{m}^{\top}\bS_{m}^{-1}\right)\Biggr]\bD_{r} \text{d}\vechop\left( \bS_{m}\right).
	\end{split}
	\label{eqn:d_F_kappa}
\end{align}
Here, we used the following properties of the commutation and duplication matrix
\begin{equation}
	\bK_{n,n} = \bK_{n}, \quad \bK_{n}\bD_{n} = \bD_{n}
\end{equation}
\begin{equation}
	(\bA \otimes \bb^{\top})\bK_{n,p} =  (\bb^{\top} \otimes \bA),
	\label{eqnA:commAB5}
\end{equation}
where $\bA \in \mathbb{R}^{m \times n}$, $\bB \in \mathbb{R}^{p \times q}$, \mbox{$\bK_{m,n} \in \mathbb{R}^{mn \times mn}$} and \mbox{$\bb \in \mathbb{R}^{p \times 1}$}. Combining the results leads to the gradient
\begin{align}
	\begin{split}
		\text{D}\bF(\bS_{m}) =&\sum_{n = 1}^{N} v_{nm} \Biggl[ -\frac{1}{2} \vecop\left(\bS_{m}^{-1}\right)^{\top} 
		\\&+  \frac{e_{0,nm}}{2}\left(\bxt_{n}^{\top}\bS_{m}^{-1} \otimes \bxt_{n}^{\top}\bS_{m}^{-1}\right) 
		\\&- \frac{e_{1,nm}}{2}\left(\bxt_{n}^{\top}\bS_{m}^{-1} \otimes  \blambda_{m}^{\top}\bS_{m}^{-1}\right) 
		\\&- \frac{e_{1,nm}}{2}\left(\blambda_{m}^{\top}\bS_{m}^{-1} \otimes \bxt_{n}^{\top}\bS_{m}^{-1}\right)
		\\& + \frac{e_{2,nm}}{2}\left( \blambda_{m}^{\top}\bS_{m}^{-1} \otimes  \blambda_{m}^{\top}\bS_{m}^{-1}\right)\Biggr]\bD_{r}.
	\end{split}
	\label{eqn:grad_S}
\end{align}
Analogously to Eq.~\eqref{eqn:grad_xi} and Eq.~\eqref{eqn:grad_lambda}, the gradient in Eq.~\eqref{eqn:grad_S} is approximated by replacing $e_{0,nm}$, $e_{1,nm}$ and $e_{2,nm}$ by their approximations, i.e. $\tilde{e}_{0,nm}$, $\tilde{e}_{1,nm}$, $\tilde{e}_{2,nm}$. Then, the approximate ML estimate follows by setting the approximate gradient to zero and solving for $\bS_{m}$:
\begin{align}
	\begin{split}
		& \sum_{n = 1}^{N} v_{nm} 
		\Biggl[ -\frac{1}{2} \vecop\left(\bS_{m}^{-1}\right)^{\top} + \frac{\tilde{e}_{0,nm}}{2} \left(\bxt_{n}^{\top}\bS_{m}^{-1} \otimes \bxt_{n}^{\top}\bS_{m}^{-1}\right) 
		\\&	- \frac{\tilde{e}_{1,nm}}{2}\left(\bxt_{n}^{\top}\bS_{m}^{-1} \otimes  \blambda_{m}^{\top}\bS_{m}^{-1}\right) \\&- \frac{\tilde{e}_{1,nm}}{2}\left(\blambda_{m}^{\top}\bS_{m}^{-1} \otimes \bxt_{n}^{\top}\bS_{m}^{-1}\right)
		\\& + \frac{\tilde{e}_{2,nm}}{2}\left( \blambda_{m}^{\top}\bS_{m}^{-1} \otimes  \blambda_{m}^{\top}\bS_{m}^{-1}\right)\Biggr]\bD_{r} \overset{!}{=} 0 
	\end{split}\notag \\
	\begin{split}
		\Rightarrow & \sum_{n = 1}^{N} v_{nm} 
		\vecop\left(\bS_{m}^{-1}\right)^{\top} \left(\bS_{m} \otimes \bS_{m}\right)  
		\\&= \sum_{n = 1}^{N} v_{nm} \Biggl[\tilde{e}_{0,nm} \left(\bxt_{n}^{\top} \otimes \bxt_{n}^{\top}\right)  - \tilde{e}_{1,nm}\left(\bxt_{n}^{\top} \otimes  \blambda_{m}^{\top}\right)
		\\&- \tilde{e}_{1,nm}\left(\blambda_{m}^{\top}\otimes \bxt_{n}^{\top}\right) + \tilde{e}_{2,nm}\left( \blambda_{m}^{\top}\otimes  \blambda_{m}^{\top}\right)\Biggr]
	\end{split}\notag \\
	\begin{split}
		\Rightarrow & \vecop\left(\bS_{m}\right) = \Biggl[\sum_{n = 1}^{N} v_{nm} \Biggl[\tilde{e}_{0,nm} \left(\bxt_{n} \otimes \bxt_{n}\right) 
		\\&- \tilde{e}_{1,nm}\left(\bxt_{n} \otimes  \blambda_{m}\right)- \tilde{e}_{1,nm}\left(\blambda_{m}\otimes \bxt_{n}\right)  
		\\&+ \tilde{e}_{2,nm}\left( \blambda_{m}\otimes  \blambda_{m}\right)\Biggr]\Biggr]\biggm/ \sum_{n = 1}^{N} v_{nm}
	\end{split}\notag \\
	\begin{split}
		\Rightarrow & \bShat_{m} = \Biggl[\sum_{n = 1}^{N} v_{nm} \Bigl(\tilde{e}_{0,nm} \bxt_{n}\bxt_{n}^{\top}  - \tilde{e}_{1,nm}\bxt_{n} \blambda_{m}^{\top} 
		\\&- \tilde{e}_{1,nm}\blambda_{m}\bxt_{n}^{\top} + \tilde{e}_{2,nm} \blambda_{m} \blambda_{m}^{\top} \Bigr)\Biggr]\biggm/ \sum_{n = 1}^{N} v_{nm}.
	\end{split}
\end{align}
Finally, we have to maximize with regard to the mixing coefficients $\gamma_{m}$. Because they have the constraint
\begin{equation}
	\sum_{m=1}^{l}\gamma_{m} = 1
\end{equation}
a Lagrange multiplier is used 
\begin{align}
	\begin{split}
		\text{d}\bF(\gamma_{m}) =& \sum_{n=1}^{N} v_{nm} \text{ d} \left(\ln(\gamma_{m})\right) + \lambda \text{d}\left(\sum_{m=1}^{l}\gamma_{m} - 1\right) 
	\end{split}\notag \\
	=& \sum_{n = 1}^{N}  \frac{v_{nm}}{\gamma_{m}} + \lambda.
\end{align}
First, we solve for $\lambda$
\begin{align}
	0 \overset{!}{=}& \sum_{n = 1}^{N} \frac{v_{nm}}{\gamma_{m}} + \lambda \notag \\
	\Rightarrow - \lambda  \sum_{m=1}^{l}\gamma_{m} =& \sum_{n = 1}^{N} \sum_{m=1}^{l} v_{nm} \notag \\
	\Rightarrow \lambda =& -N
\end{align}
and, after the elimination of $\lambda$, we find
\begin{align}
	\hat{\gamma}_{m} =& \frac{1}{N} \sum_{n = 1}^{N} v_{nm}.
\end{align}

\section*{Acknowledgment}
Christian A. Schroth is supported by the DFG Project Number 431431951. The work of Michael Muma has been funded by the LOEWE initiative (Hesse, Germany) within the emergenCITY centre and is supported by the ‘Athene Young Investigator Programme’ of Technische Universität Darmstadt, Hesse, Germany.

\renewcommand*{\bibfont}{\footnotesize}
\printbibliography

\end{document}